\documentclass[sn-mathphys, twocolumn]{sn-jnl}

\usepackage{graphicx}%
\usepackage{multirow}%
\usepackage{amsmath,amssymb,amsfonts}%
\usepackage{amsthm}%
\usepackage{mathrsfs}%
\usepackage[title]{appendix}%
\usepackage{xcolor}%
\usepackage{textcomp}%
\usepackage{manyfoot}%
\usepackage{booktabs}%
\usepackage{algorithm}%
\usepackage{algorithmicx}%
\usepackage{algpseudocode}%
\usepackage{listings}%

\usepackage{geometry}
\begin{document}

\title{Experimental observations of microturbulence-suppressed parallel heat conduction in a weakly collisional, high-$\beta$ plasma}

\author*[1]{\fnm{T. A.} \sur{Vincent}}\email{thomas.vincent@physics.ox.ac.uk}

\author[2]{\fnm{P.} \sur{Ariyathilaka}}
\author[3]{\fnm{L.} \sur{Creaser}}
\author[1,3]{\fnm{C.} \sur{Danson}}
\author[1]{\fnm{R.} \sur{Davies}}
\author[4]{\fnm{D.} \sur{Lamb}}
\author[5]{\fnm{J.} \sur{Meinecke}}
\author[6]{\fnm{C.A.J.} \sur{Palmer}}
\author[3]{\fnm{S.} \sur{Pitt}}
\author[7]{\fnm{H.} \sur{Poole}}
\author[2]{\fnm{C.} \sur{Spindloe}}
\author[3]{\fnm{P.} \sur{Thomas}}
\author[8]{\fnm{E. R.} \sur{Tubman}}
\author[3]{\fnm{L.} \sur{Wilson}}
\author[3]{\fnm{W.} \sur{Garbett}}
\author[1]{\fnm{G.} \sur{Gregori}}
\author[7]{\fnm{P.} \sur{Tzeferacos}}
\author[3]{\fnm{T.} \sur{Hodge}}

\author*[1]{\fnm{A. F. A.} \sur{Bott}}\email{archie.bott@physics.ox.ac.uk}

\affil*[1]{\orgdiv{Department of Physics}, \orgname{University of Oxford}, \orgaddress{\street{Parks Rd}, \city{Oxford}, \postcode{OX1 3PU}, \country{UK}}}
\affil[2]{\orgdiv{Central Laser Facility}, \orgname{Rutherford Appleton Laboratory}, \orgaddress{\city{Didcot}, \postcode{OX11 0OX}, \country{UK}}}
\affil[3]{\orgname{AWE}, \orgaddress{\street{Aldermaston}, \city{Reading}, \postcode{RG7 4PR}, \country{UK}}}
\affil[4]{\orgdiv{Department of Astronomy and Astrophysics}, \orgname{University of Chicago}, \orgaddress{\city{Chicago}, \postcode{IL 60637}, \country{USA}}}
\affil[5]{\orgdiv{Department of Physics}, \orgname{Gettysburg College}, \orgaddress{\city{Gettysburg}, \postcode{PA 17235}, \country{USA}}}
\affil[6]{\orgdiv{School of Mathematics and Physics}, \orgname{Queen's University Belfast}, \orgaddress{\city{Belfast}, \postcode{BT7 1NN}, \country{UK}}}
\affil[7]{\orgdiv{Department of Physics and Astronomy}, \orgname{University of Rochester}, \orgaddress{\city{Rochester}, \postcode{NY 14627}, \country{USA}}}
\affil[8]{\orgdiv{Department of Nuclear Engineering}, \orgname{University of California Berkeley}, \orgaddress{\city{California}, \postcode{CA 94720}, \country{USA}}}

\abstract{Classical theories of heat conduction in magnetized plasma predict that thermal transport, mediated by Coulomb collisions of electrons, occurs predominantly along magnetic field lines. Recent theoretical and computational studies challenge this description for weakly collisional, magnetized plasmas in which thermal pressure dominates magnetic pressure (so-called high-$\beta$ plasmas). Such plasmas, which include the intracluster medium of galaxy clusters and inertial-confinement-fusion hot spots, are thought to be susceptible to kinetic-scale microinstabilities that can suppress parallel heat conduction. Revised theories of heat conduction accounting for these instabilities have been proposed, but experimental data with which to benchmark them are lacking. Here, we report experiments at the Orion laser facility in which the temporal evolution of temperature of a high-$\beta$, weakly collisional plasma is sensitive primarily to the thermal conductivity along its initially quasi-laminar magnetic field. We characterize the temperature, density and magnetic field using x-ray spectroscopy and imaging, and proton imaging. Once stochastic magnetic fluctuations develop, the measured temperature evolution requires suppression of thermal conduction by at least an order of magnitude relative to classical predictions, providing the first laboratory evidence linking magnetic microturbulence with suppression of heat conduction in a high-$\beta$ plasma.}

\maketitle

The formation and evolution of many astrophysical and laboratory plasmas are strongly influenced by how they conduct heat.  
For example, in the intracluster medium (ICM) of galaxy clusters, the efficiency of heat conduction has important implications for long-standing puzzles such as the cooling-flow problem~\cite{Fabian_1994} and the persistence of sharp temperature structures associated with cold fronts~\cite{Ascasibar_2006}.  
In inertial-confinement fusion (ICF) plasmas, heat conduction from the hot fuel is a major energy-loss mechanism and can impede ignition~\cite{abu-shawareb_2022, meezan_2020}. 
Accurate models of heat transport across a range of plasma conditions are therefore crucial for addressing key challenges in contemporary plasma physics.     

The first theories of heat conduction in plasmas were developed in the mid-twentieth century. These so-called classical transport theories assume that the electron mean free path $\lambda_e$ due to Coulomb collisions is much smaller than the characteristic temperature-gradient length scale $L_{\rm T} \equiv T_e/|\nabla T_e|$, where $T_e$ is the electron temperature. Under this assumption, heat transport is diffusive and carried predominantly by electrons. In an unmagnetized plasma, the electron heat flux follows Fourier's law, $\boldsymbol{q} = - \kappa_{\mathrm{S}} \nabla T_e$, where the Spitzer-H\"arm thermal conductivity is $\kappa_{\mathrm{S}} = \mathcal{A}(Z)n_e\lambda_e^2 \nu_e$~\cite{spitzer_1953}. Here, $n_e$ the electron number density, $\nu_e$ the electron collision frequency, and $\mathcal{A}(Z)$ is an order-unity constant that depends on the ion charge $Z$. In magnetized plasmas in which the Larmor radius $\rho_e$ of electrons is much smaller than $\lambda_e$, heat conduction becomes highly anisotropic: the heat flux $q_{e\|} = -\kappa_{\mathrm{S}} \nabla_{\|} T_e$ along magnetic field lines continues to obey the Spitzer-H\"arm model, whereas the perpendicular conductivity is suppressed by a factor ${\sim}\rho_e^2/\lambda_e^2$~\cite{braginskii_1965}. 

An emerging body of observational and experimental evidence suggests that classical transport models can substantially overestimate heat conduction in magnetized and weakly collisional ($\rho_e \ll \lambda_e \ll L_{\rm T}$) plasmas in which the thermal pressure greatly exceeds the magnetic pressure. Such plasmas have a high plasma beta,  $\beta \equiv 8\pi p/B^{2} \gg 1$, where $p$ and $B$ are the thermal pressure and macroscopic magnetic-field strength, respectively. In the ICM, observations of long-lived temperature structure imply that effective thermal conduction can be suppressed by orders of magnitude relative to the Spitzer-H\"arm value~\cite{richard-laferriere_2023}.  High-resolution x-ray spectroscopy with XRISM provides further support for inefficient thermal transport in cluster cores: the presence of strong temperature structure despite the low levels of turbulent plasma motion inferred from line broadening is challenging to reconcile with estimates based on classical transport~\cite{XRISM_2025}. In the laboratory, a recent laser-plasma experiment at the National Ignition Facility (NIF)~\cite{moses_2008} provided evidence of strongly suppressed heat conduction in a turbulent, magnetized high-$\beta$ plasma, with the inferred conductivity being around two orders of magnitude below the unmagnetized Spitzer-H\"arm value~\cite{meinecke_2022}. The apparent breakdown of classical transport theory poses a significant challenge for modeling a range of weakly collisional, high-$\beta$ plasma environments, including the ICM and ICF hot spots~\cite{abu-shawareb_2022}, the warm intergalactic medium (IGM)~\cite{nicastro_2008}, and laser-produced plasmas~\cite{tzeferacos_2018, bott_pnas_2021}. 

A plausible explanation for these observations is that sufficiently high-$\beta$ plasmas are generically susceptible to plasma microinstabilities on scales below the Coulomb mean free path~\cite{bott_2024}. The transport of heat down a temperature gradient requires a small anisotropic distortion of the electron distribution function; when $\beta \gtrsim L_{\rm T}/\lambda_e$, this distortion can destabilize electromagnetic fluctuations that subsequently scatter or trap electrons, thereby modifying their distribution function and suppressing thermal conduction. One simple model of this phenomenon, independent of the specific instability, was proposed by Ryutov et al.~\cite{ryutov_1999}; if destabilized magnetic fluctuations have a characteristic scale $\ell_{\rm B}$ and attain amplitudes comparable to that of the macroscopic magnetic field, scattering of heat-carrying electrons yields an effective conductivity $\kappa_{\rm R} \approx (\ell_{\rm B}/\lambda_e)\kappa_{\rm S} $. 

More specifically, over the past few decades, the whistler heat-flux instability (WHFI) has emerged as a leading candidate for suppressing heat conduction~\cite{levison_eichler_1992}. The WHFI draws free energy from the anisotropic electron distribution function associated with a temperature gradient along the macroscopic magnetic field, driving the growth of whistler waves that scatter heat-carrying electrons as they saturate~\cite{pistinner_eichler_1998}. Analytical~\cite{drake_2021,lopez_2025} and numerical~\cite{komarov_2018, roberg-clarke_2016, roberg-clarke_2018, yerger_2025,choudhury_2026a} studies have shown that 
this process suppresses the parallel heat flux and therefore the effective parallel thermal conductivity. These studies broadly agree that, in the strongly whistler-regulated regime, the saturated heat flux becomes independent of the magnitude of the temperature gradient and instead scales inversely with the electron plasma beta $\beta_{e} \equiv 8 \pi p_e/B^2$, where $p_e$ is the electron thermal pressure. For example, Yerger et al.~\cite{yerger_2025} find a strongly suppressed parallel heat flux $q_{e\|} \approx 1.2 n_e T_e v_{\mathrm{th}e}/\beta_e$, where $v_{\mathrm{th}e} = \sqrt{2 T_e/m_e}$ is the electron thermal speed and $m_e$ is the electron mass, corresponding to an effective thermal conductivity of $\kappa_{\rm Y} \approx 0.8 [\beta_{e}\lambda_{e}/L_{T}]^{-1} \kappa_{\rm S}$, in the limit $\beta_{e}\lambda_{e}/L_{T} \gg 1$. 

 Laser-plasma experiments provide a complementary means of testing whether kinetic microinstabilities can suppress thermal transport under controlled conditions. This capability is particularly important because astronomical observations such as the recent XRISM measurements constrain the macroscopic consequences of suppressed transport, but cannot directly resolve the electron-scale processes responsible for it. Experimental studies have previously explored modifications to thermal conductivity arising from non-local effects~\cite{henchen_2018} and the presence of magnetic fields~\cite{froula_2007}, but no experiments have investigated microinstability-mediated transport in a kinetically unstable, high-$\beta$ plasma. Such experiments can also test theoretical predictions in regimes that remain challenging to capture with simulations. No numerical studies of the WHFI has yet incorporated Coulomb collisionality, instead treating the instability as a collisionless phenomenon, while analytical studies have required simplifying assumptions to remain tractable. To reduce computational cost, numerical studies have also adopted idealized configurations, such as two-dimensional domains with linear temperature gradients maintained by thermal reservoirs. Capturing the three-dimensional interplay between macroscopic dynamics and kinetic microinstabilities is intrinsically multiscale and remains computationally demanding.

\begin{figure}
    \centering
    \includegraphics[width=1\linewidth]{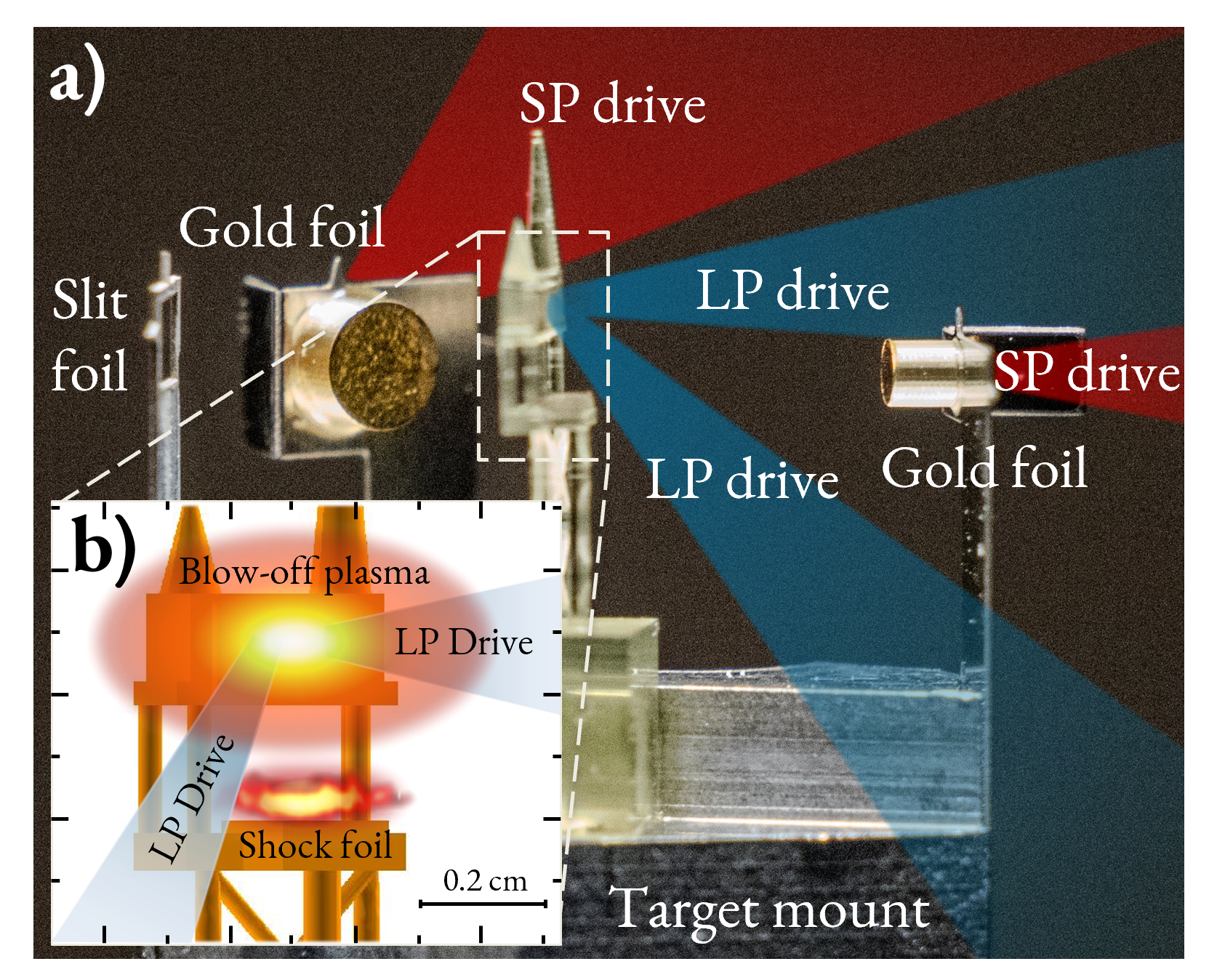}
    \caption{\textbf{Target set-up for shocked plasma production.} \textbf{a} Photograph of target, overlaid with long pulse (LP) drive beams (blue), and short pulse (SP) beams for proton generation (red). Each LP irradiates the target with a $0.3~\text{kJ}$, $1~\text{ns}$-duration, $351~\text{nm}$-wavelength approximate super-Gaussian beam, with a $0.23~\text{mm}$ focal spot. Each SP irradiates the target with a $0.3~\text{kJ}$, $1~\text{ps}$-duration, $1054~\text{nm}$-wavelength beam, with an approximate $7.7~\mu\text{m}$ focal spot determined from the full-width half-maximum (FWHM) of the short-pulse driven x-ray emission. \textbf{b} Schematic of the target overlaid with x-ray emission data from the shocked plasma.}
    \label{fig:target_diagram}
\end{figure}

In this work, we present results from experiments conducted at the Orion laser facility~\cite{hopps_2015}, providing the first experimental evidence of magnetized kinetic microturbulence accompanied by strongly suppressed parallel thermal conduction in a weakly collisional, high-$\beta$ plasma. Fig. \ref{fig:target_diagram}a shows the experimental target configuration. 
Two co-timed long-pulse drive beams irradiate the target's drive foil, separated horizontally by $L_{s}=0.5~\text{mm}$, producing a front-side blow-off plasma~\cite{amoruso_1999}. The resulting supersonic plasma plume expands radially from the irradiated surface with an approximately oblate hemispherical profile and subsequently shocks onto a secondary (shock) foil, compressing and rapidly heating the plasma~\cite{colvin_2013}. During ablation of the drive foil, magnetic fields are generated via the Biermann battery mechanism~\cite{biermann_1950, tubman_2021, pearcy_2025}. The blow-off plasma flow has a large magnetic Reynolds number $\text{Rm} = u_{\text{plume}}L_{s}/\eta\approx 2\times 10^{4}$, where $u_{\text{plume}}\approx 9 \times 10^{7}~\text{cm s}^{-1}$ is the flow velocity, and $\eta\approx300~\text{cm}^{2}~\text{s}^{-1}$ the plasma resistivity (see Methods). Under these conditions, the Biermann-generated fields are approximately frozen into the flow in accordance with Alfvén's theorem~\cite{alfven_1943}, and are subsequently compressed at the shock. Fig. \ref{fig:target_diagram}b illustrates the resulting quasi-planar volume of shocked plasma. This platform enables the reproducible generation of a weakly collisional, high-$\beta$ plasma with well-defined geometry and magnetic-field structure, suitable for characterising heat conduction. Diagnostics include proton imaging to infer magnetic field strength and morphology, as well as filtered soft x-ray self-emission and x-ray emission spectroscopy for temperature and density inference. The platform's design~\cite{vincent_2026} was informed by a series of three-dimensional radiation magnetohydrodynamic simulations performed using FLASH~\cite{fryxell_2000, tzeferacos_2015}, with post-processing carried out using SPECT3D~\cite{macfarlane_2006,macfarlane_2007} and PrismSPECT~\cite{golovkin_2025} to optimize diagnostic configuration and target parameters.

\begin{figure*}
    \centering
    \includegraphics[width=1\linewidth]{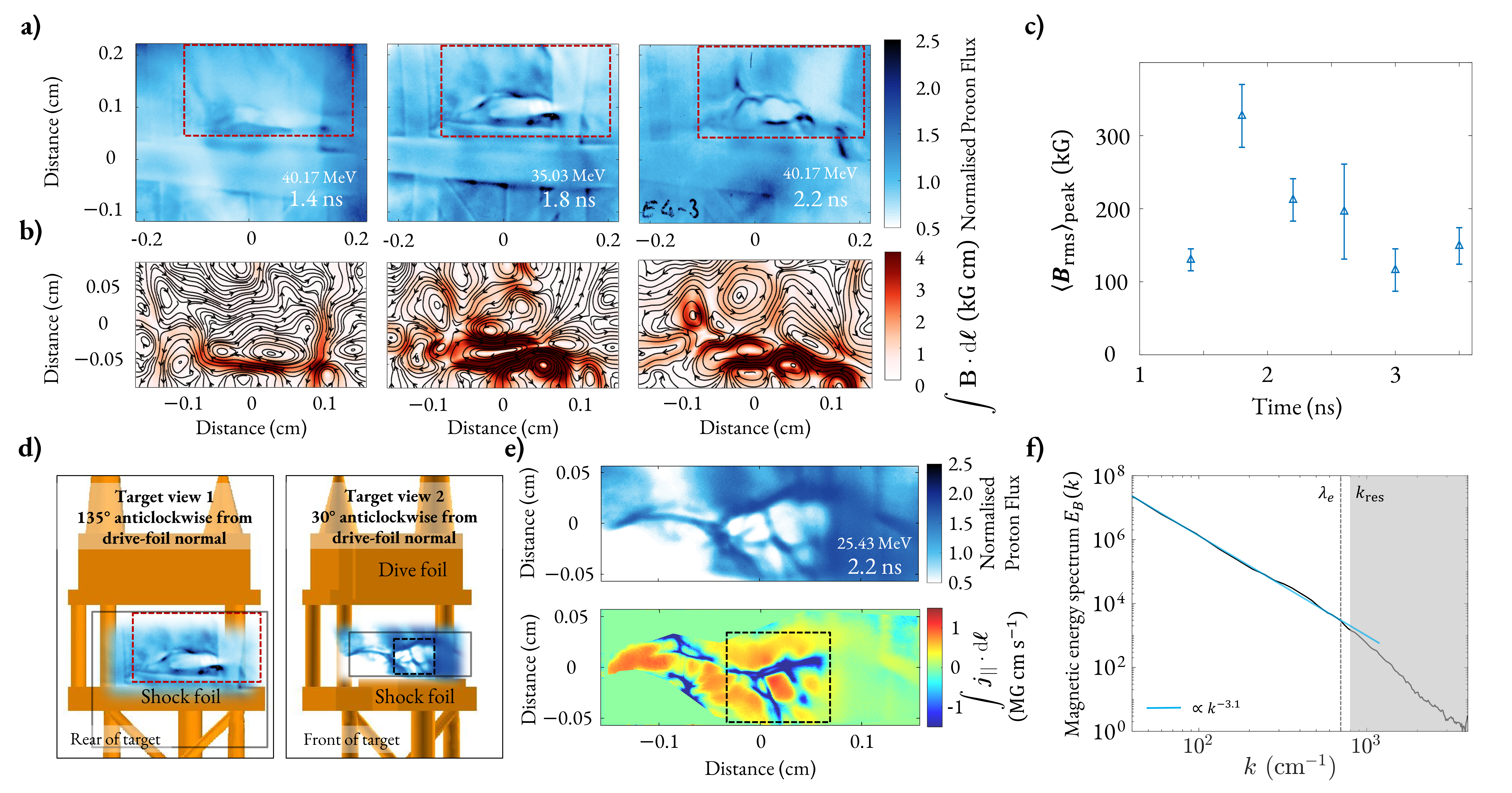}
    \caption{\textbf{Proton-fluence data and magnetic-field reconstructions.} \textbf{a} Normalised proton fluence images of the shocked plasma region between 1.4 and 2.2 ns, as seen through target view 1, shown in (d). The fluence images were captured at a distance from the target such that they have a $\times23$ magnification. Each panel is labelled with the proton energy and time after the start of the long-pulse drive. The red dashed box outlines the region over which the magnetic field reconstruction is performed. \textbf{b} Path-integrated magnetic-field reconstructions of the corresponding fluence images shown in (a). \textbf{c} Evolution of the peak root-mean-square magnetic field, with error bars indicating one standard deviation from the mean (Methods). \textbf{d}. Schematic of target from two viewing angles; target view 1 and view 2 corresponding to $30^{\circ}$ and $135^{\circ}$ anticlockwise from the drive target normal, respectively. Each target view is overlaid with proton fluence data, from (a) and (e), to illustrate the signals' respective position relative to target structures. The black box surrounding the fluence signal corresponds to the extent of the data shown in the rest of the figure, the red dashed box shows the reconstruction region, and the grey dashed box shows the region used for the power spectrum analysis in (f). \textbf{e} Proton-imaging data at 2.2 ns, as seen through target view 2. The top panel shows the normalized proton fluence; the bottom panel shows the reconstructed path-integrated electron current density parallel to the line of sight over the indicated region. \textbf{f} Path-integrated magnetic-energy spectrum corresponding to the data shown in (e). A super-Gaussian windowing function is applied to the current map, setting the windowed current to zero outside of the gray dashed box. The solid pale-blue line indicates a $k^{-3.1}$ spectrum, while the gray dashed line marks the electron mean-free path scale at $2.2~\text{ns}$. The gray shaded region at $k\geq k_{\text{res}}$ denotes the diagnostic-resolution cut-off inferred from the power spectrum of the proton-fluence image in (e). The source size sets a maximum bound on the diagnostic resolution, stated in the main text.}
    \label{fig:PRAD_data}
\end{figure*}

Magnetic fields are characterized using proton radiography (PRAD) \cite{kugland_2012, bott_2017, schaeffer_2023}, with imaging performed along two lines of sight at $30^{\circ}$ and $135^{\circ}$ anticlockwise, respectively, relative to the normal of the long-pulse drive foil (see Fig.~\ref{fig:PRAD_data}d). 
 The proton beams are generated via target-normal sheath acceleration (TNSA)\cite{macchi_2013} by irradiating two gold foils with Orion's petawatt short-pulse beams (see Methods). The protons then pass through the plasma, and their fluence is detected on radiochromatic film (see Supplementary Information for further details). Fig.\ref{fig:PRAD_data}a shows proton-fluence images obtained using $35$-$40~\text{MeV}$ protons over the interval $1.4$-$2.2~\text{ns}$ after the start of the long-pulse drive, viewed along the same line-of-sight indicated in the left panel of Fig.\ref{fig:PRAD_data}d. 
The region enclosed by the red dashed box corresponds to the plasma of interest, with its vertical centreline aligned with the left wing of the plasma. At $1.4~\text{ns}$, laminar fluence variations appear directly above the shadow of the shock foil, while a localized low-fluence region farther above indicates a net downward deflection of protons. By $1.8~\text{ns}$, smaller-scale structures begin to emerge on either side of this region, evolving into larger, stochastic features by $2.2~\text{ns}$. These structures coincide spatially with regions of the strongest temperature gradients.

Path-integrated magnetic field maps (Fig.~\ref{fig:PRAD_data}b) can be reconstructed from the proton fluence images under the assumptions of small proton deflections and the absence of caustics~\cite{kugland_2012, bott_2017}. By analyzing the highest-energy protons produced by the short-pulse lasers that still give a full image, caustic formation can be avoided~\cite{bott_prl_2021}. These reconstructions corroborate the interpretation of the fluence profiles, revealing a smooth, regular magnetic structure in the shocked plasma at early times and the development of stochastic features by $2.2~\text{ns}$. Fig.~\ref{fig:PRAD_data}c shows the temporal evolution of the peak root-mean-square magnetic field $\langle \boldsymbol{B}_{\mathrm{rms}}\rangle_{\text{peak}}$, calculated assuming a constant path length $\ell_{\text{p}}\approx0.028~\text{cm}$, corresponding to the approximate width of the interaction region (Methods). The initial increase in magnetic-field strength between $1.4$ and $1.8~\text{ns}$ is attributed to field advection and compression of the flux-frozen magnetic fields into the shocked region, followed by dissipation at later times as the plasma expands. From this analysis, we obtain $\langle \boldsymbol{B}_{\mathrm{rms}} \rangle_{\mathrm{peak}} = 212 \pm 29~\text{kG}$ at $2.2~\text{ns}$.

Fig.~\ref{fig:PRAD_data}e shows the proton fluence and line-of-sight-integrated electron-current component parallel to the proton trajectories at $2.2~\text{ns}$ after the start of the long-pulse drive, for the line-of-sight to that shown in the left panel of Fig.~\ref{fig:PRAD_data}d, target view 2. In the fluence data, stochastic structures are visible between $-0.05$ and $0.05~\text{cm}$, although they are obscured along the lines of sight shown in the left panel of Fig.~\ref{fig:PRAD_data}d, target view 1. Owing to the geometry of target view 2, the macroscopic magnetic fields defocus the proton beam, locally enhancing the image magnification in one projection and increasing the apparent size of these stochastic structures. In Fig.~\ref{fig:PRAD_data}a, target view 1, the opposite occurs -- the proton beam is focused -- and the structures are less apparent~\cite{kugland_2012}. This geometrical distortion is removed automatically as part of the reconstruction algorithm, revealing current structures across a broad range of scales~\cite{bott_2017}. More quantitatively, Fig.~\ref{fig:PRAD_data}f shows the line-of-sight-integrated magnetic-energy spectrum corresponding to the data in Fig.~\ref{fig:PRAD_data}e. The spectrum approximately follows a $k^{-3.1}$ power-law down to scales approaching the fluence-resolution cutoff, illustrated by the gray shaded region for $k\gtrsim k_{\text{res}}$. The characteristic scale of the smallest resolvable stochastic structures is therefore $\ell_{\mathrm{B,min}}\approx\pi/k_{\text{res}}\approx 40~\mu\text{m}$.

\begin{figure}
    \centering
    \includegraphics[width=1\linewidth]{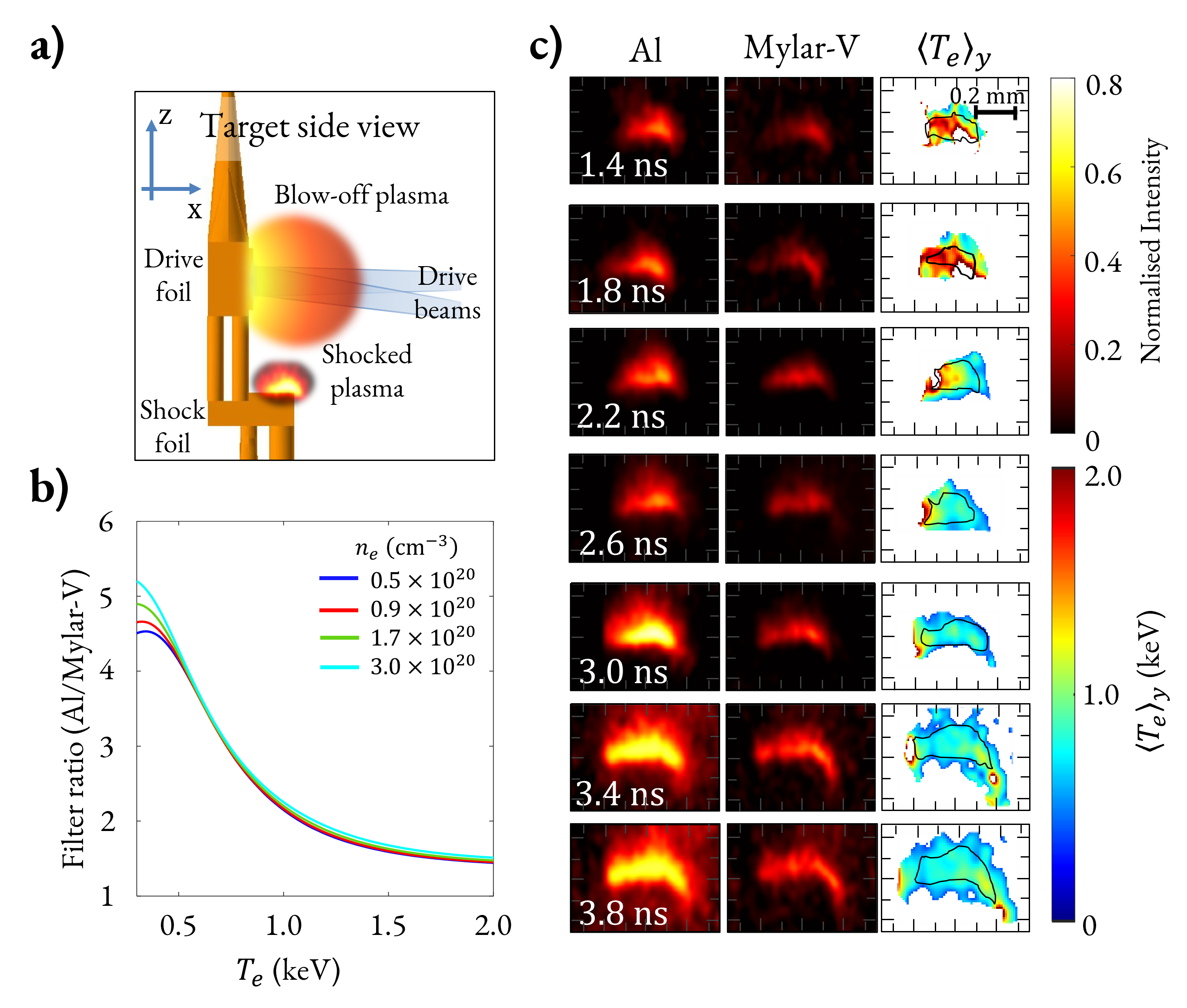}
    \caption{\textbf{X-ray self-emission data and inferred temperature maps.} \textbf{a} Schematic of the target overlaid with x-ray data, illustrating the side-on view of the shocked plasma. \textbf{b} Al-to-Mylar-V filter ratio-curves at fixed $n_{e}=0.5$-$3.0\times10^{20}~\text{cm}^{-3}$. \textbf{c} .Evolution of the x-ray self-emission filtered by 0.544 $\mu$m Al (left) and 0.9 $\mu$m Mylar + 0.416 $\mu$m V (center), with corresponding filter-ratio-inferred temperature maps (right). The black contour encloses the region with emission exceeding $30\%$ of the peak mylar-V signal. Owing to the diagnostic sensitivity in the relevant temperature-density regime~\cite{vincent_2026}, this contour defines the region over which the average temperatures shown in Fig.\ref{fig:temperature_evolution_and_xRay_spectrum}b are calculated. The x-ray self-emission was recorded using a gated x-ray detector (GXD) coupled to a pinhole array and micro-channel plate (MCP)~\cite{rochau_2006}, providing five-times magnification, $9.5~\mu\text{m}/\text{pix}$ resolution, and $200~\text{ps}$ intra-strip sweep. The MCP consists of a $3\times4$ pinhole array, with $10~\mu\text{m}$ pinholes, arranged into two pairs of synchronised strips with a $400~\text{ps}$ delay between pairs. See Methods for further details of ratio-curve analysis, and required signal flat-fielding procedure relating to the intra-strip sweep.}
    \label{fig:gxd_temp_map_data}
\end{figure}

Soft x-ray self-emission imaging is used to characterize the dynamics of the shocked plasma as well as its electron temperature $T_{e}$ and number density $n_e$. Fig.~\ref{fig:gxd_temp_map_data}c shows spatially resolved x-ray images of the plasma obtained through two broadband filters: aluminum (Al) and mylar-vanadium (Mylar-V). 
The line of sight associated with these images is illustrated in Fig.~\ref{fig:gxd_temp_map_data}a. Qualitatively, emission from the shocked plasma first appears $1.4~\text{ns}$ after the start of the long-pulse drive -- concurrently with the emergence of a compressed magnetic-field structure -- and persists until at least $3.8~\text{ns}$. Temperature maps are obtained  by taking the ratio of co-timed images from the same shot, yielding two-dimensional, line-of-sight-integrated, density-weighted temperature maps, $\langle T_e \rangle_y$ \cite{meinecke_2022, valenzuela-villaseca_2024}, using the $n_{e}=0.9\times 10^{20}~\text{cm}^{-3}$ filter-ratio curve shown in Fig.~\ref{fig:gxd_temp_map_data}b. We attribute the apparent inhomogeneities in these temperature maps to photon-counting statistics arising from the low signal in the Mylar-V-filtered images, rather than to intrinsic plasma structure, consistent with previous observations~\cite{valenzuela-villaseca_2024}. Validation of this diagnostic configuration using synthetic FLASH data, including the filter selection and line of sight use, is described in Vincent et al.~\cite{vincent_2026}. 

The electron density of the shocked plasma is inferred from the relative brightness of the peak emission from the ablation plume and the spatially averaged emission within the contour displayed in Fig.~\ref{fig:gxd_temp_map_data}c, as measured through the Al filter (Methods). This emissivity calibration yields $n_{e} = (1.0\pm0.6)\times10^{20}~\text{cm}^{-3}$ at $2.2$ ns after the start of the long-pulse drive. This inferred density is independently consistent with the observed drop in x-ray emission between the shocked plasma and the inflowing ablation plume: the measured shock-transition width, $\ell_{\rm sh} \approx 4 \times 10^{-3} \; \mathrm{cm}$ (Extended Data Fig.~\ref{fig:gxd_transmission}a), is comparable to the Coulomb mean free path $\lambda_{\mathrm{C},\mathrm{plume}} \approx 4 \times 10^{-3} \; \mathrm{cm}$ of the carbon ions in the ablation plume plasma streaming through the shocked plasma for $n_{e} \approx 10^{20}~\text{cm}^{-3}$ (see Methods). The inferred density is also consistent with measurements of Cl-doped CH plasmas produced under similar experimental conditions at the same laser facility~\cite{tubman_2021}. 

\begin{figure*}
    \centering
    \includegraphics[width=1\linewidth]{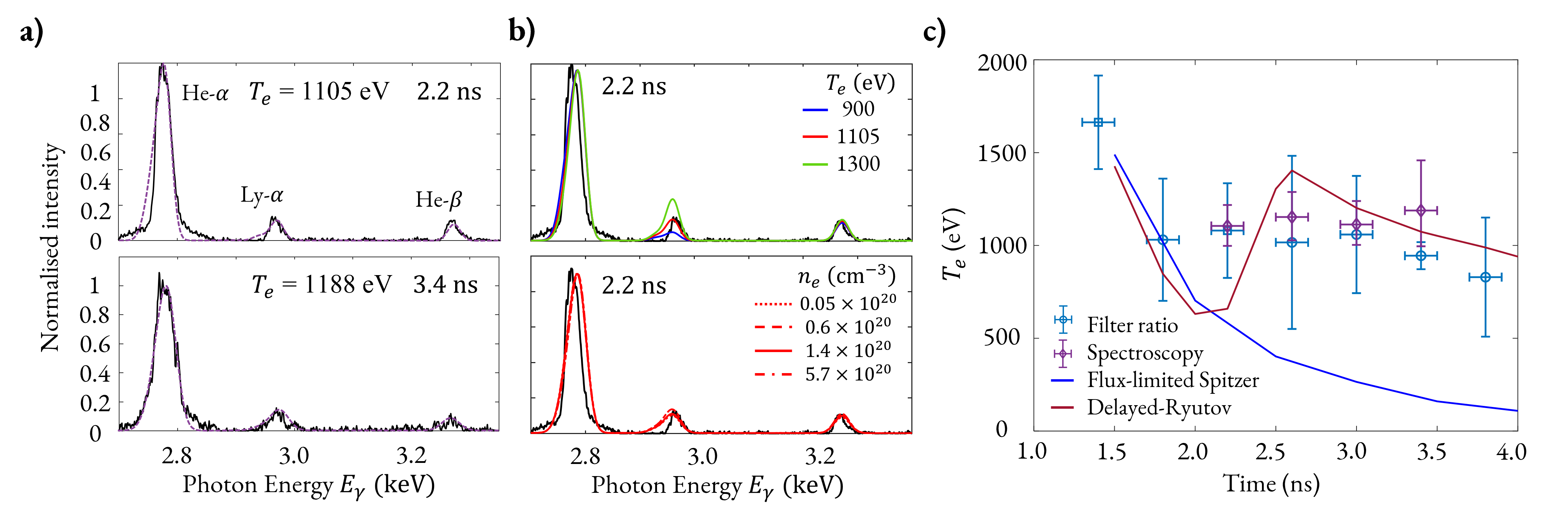}
    \caption{\textbf{Spectral fitting and plasma temperature evolution.} \textbf{a} Self-emission x-ray spectra measured between $2.2$ and $3.4~\text{ns}$. The spectral window contains the He-$\alpha$, Ly-$\alpha$, and He-$\beta$ spectral peaks, labelled in the $2.2~\text{ns}$ panel. The dashed purple lines show single-cell simulated spectra at the average temperature inferred from the Ly-$\alpha$/He-$\alpha$ and Ly-$\alpha$/He-$\beta$ line ratios, at fixed $n_{e}=1.4\times10^{20}~\text{cm}^{-3}$. \textbf{b} Synthetic single-cell fits have been overlaid onto the experimental data at {2.2~\text{ns}} for a range of $T_{e}$ and $n_{e}$. The top panel shows fits at fixed $n_{e}=1.4\times 10^{20}~\text{cm}^{-3}$, and varying $T_{e}=900$-$1300~\text{eV}$. The bottom panel shows fits at fixed $T_{e}=1105~\text{eV}$, and varying $n_{e}=0.05$-$5.7\times10^{20}~\text{cm}^{-3}$. These series of fits illustrate the spectrums sensitivity to temperature change, while being insensitive to changes in density. Although not shown here, the spectra are also weakly insensitive to the size of the emitting region. \textbf{c} Temporal evolution of the plasma's electron temperature. Temperatures inferred from the x-ray self-emission ratio-map analysis and x-ray spectroscopy are indicated by the circles and diamonds, respectively. Vertical error bars for the x-ray self-emission inferred $T_{e}$ show one standard deviation from the mean, calculated from repeat measurements. The two points indicated by square markers, insufficient repeat measurements were obtained; their vertical uncertainties therefore correspond to one standard deviation from the mean within the spatially averaged region. The vertical uncertainties for the x-ray spectroscopy inferred $T_{e}$ are obtained from $90\%$ confidence intervals on the spectral line fits (see Supplementary Information for details on background removal). The horizontal error bars for the x-ray self-emission, and spectroscopy span the $200~\text{ps}$ intra-strip sweep the detector used when measuring the emission. The solid navy-blue line indicates the temperature evolution inferred from FLASH simulation data using flux-limited Spitzer-H\'arm as the conductivity model. The solid maroon line indicates temperatures inferred by synthetic, post-processed, spectroscopy applied to the delayed-Ryutov FLASH simulation. Further details of the synthetic FLASH analysis are provided in the Supplementary Information.}
    \label{fig:temperature_evolution_and_xRay_spectrum}
\end{figure*}

X-ray emission spectroscopy provides a second, independent diagnostic of temperature. The drive foil is composed of parylene C (PyC), containing $50\%$ carbon (C), $43.75\%$ hydrogen (H), and $6.25\%$ chlorine (Cl) by atomic fraction, enabling observation of He-like and H-like Cl spectral lines in the shocked plasma after $2.0$ ns along the same line-of-sight as the x-ray self-emission imaging. The spectrometer employs an RbAP crystal covering the $2.5$–$5.0~\text{keV}$ spectral range. The plasma is optically thin to these spectral lines once broadening is taken into account (see Extended Data Fig.~\ref{fig:opacity_and_line_ratios}). Fig.~\ref{fig:temperature_evolution_and_xRay_spectrum}a shows spectra measured between $2.2$ and $3.4~\text{ns}$ after the start of the long-pulse drive, containing the $1\text{s}2-1\text{s}2\text{p}$ $1\text{S}-1\text{P}$ (He-$\alpha$), $1\text{s}-2\text{p}$ (Ly-$\alpha$), and $1\text{s}2-1\text{s}3\text{p}$ $1\text{S}-1\text{P}$ (He-$\beta$) transitions. These lines are identified using atomic transition data from PrOpacEOS~\cite{macfarlane_2006}, see Extended Data Fig.~\ref{fig:spectral_line_identification} for full list of identified emission lines. Electron temperatures were inferred from fits to the Ly-$\alpha$, He-$\alpha$, and He-$\beta$ lines. The resulting line-ratio temperatures are validated by comparison with single-cell synthetic spectra at fixed density and temperature. Examples of these fits are shown by the purple dashed lines in Fig.~\ref{fig:temperature_evolution_and_xRay_spectrum}a. The sensitivity of these synthetic fits to temperature, and density variations are shown in Fig.~\ref{fig:temperature_evolution_and_xRay_spectrum}b.
Uncertainties are derived from the $90\%$ confidence intervals of the individual spectral-line fits before calculating the line ratios.  See Methods for further details on spectral line identification, line fitting, and temperature inference from fitted line ratios.

Fig.~\ref{fig:temperature_evolution_and_xRay_spectrum}c shows the experimentally inferred temporal evolution of $T_{e}$ within the shocked plasma region, together with post-processed synthetic FLASH data. Our preliminary simulations employ flux-limited Spitzer-H\"arm conductivity, capped at $6\%$ of the free-streaming heat-flux limit. Temperatures inferred from filter-ratio maps and x-ray spectroscopy agree within their respective uncertainties, with the spectroscopic measurements providing more precise estimates of $T_{e}$. The experimental temperatures agree well with the Spitzer-based simulation up to $2.0~\text{ns}$ after the start of the long-pulse drive. At later times, however, a significant deviation emerges; the experimental temperatures remains approximately constant at $\langle T_{e}\rangle_{y}\approx 1.1~\text{keV}$ until $3.8~\text{ns}$. Reproducing this lack of cooling requires an effective conductivity $\kappa_{\text{eff}}$ that is at least an order of magnitude smaller than $\kappa_{\text{S}}$ in our synthetic modeling. An analysis of the thermal-energy equation, based on synthetic FLASH data, confirms that under such suppression the dominant energy sources and sinks approximately balance (Methods). The primary energy sources are injection from the ablation plume into the shocked region and ion–electron equilibration. During shock heating, ions are preferentially heated relative to electrons~\cite{langendorf_2018}, and subsequently transfer thermal energy to the electrons at later times. The dominant energy sinks are thermal conduction, adiabatic cooling due to expansion, and radiative losses primarily via line emission.  

Using the measured plasma parameters at the onset of large stochastic magnetic fluctuations, we assess the predicted susceptibility of the shocked plasma to the WHFI (expressions used to calculate these parameters are provided in the Extended Data Table \ref{table:equations_for_calcs}). A primary requirement for the development of the instability is a substantial temperature gradient along the magnetic field. In this experiment, the dominant temperature gradients extend from the center of the shock foil into the cooler wings as the plasma expands. Early-time proton imaging (Fig.~\ref{fig:PRAD_data}b) confirms that the magnetic field is initially laminar and oriented along the major axis of the shock foil, and is therefore approximately parallel to the temperature gradients. The plasma parameters, $\beta_{e} \approx 100$, $L_{T} \approx 600~\mu\text{m}$, and $\lambda_{e} \approx 40~\mu\text{m}$, satisfy the condition $\lambda_{e}/L_{T} \gtrsim \beta_{e}^{-1}$ for instability onset. The characteristic timescale of linear growth of the WHFI, $\tau_{\text{WHFI}} \approx 4 \times 10^{-11}~\text{s}$~\cite{bott_2024}, is shorter than the Spitzer thermal diffusion timescale, $\tau_{\kappa_{\rm S}} \equiv 2 n_{e} L_{\rm T}^2/(3 \kappa_{\rm{S}}) \approx 2 \times 10^{-10}~\text{s}$, indicating that the instability can undergo several e-folding times before classical conduction substantially cools the plasma. The experimental plasma is therefore expected to be susceptible to the WHFI. Although the stochastic magnetic-field fluctuations in our experiment cannot be identified directly as whistler modes, the inferred minimum resolved fluctuation length scale $\ell_{\mathrm{B,min}} \lesssim \lambda_{e}$, indicates that the fluctuations extend to scales comparable to, or smaller than, the electron mean free path see Fig.~\ref{fig:PRAD_data}f. The observed magnetic structure therefore extend down to the collisionless range, consistent with a kinetic origin.

Having established that the conditions expected to support WHFI growth, we assess the reduction in the thermal conductivity predicted by current theories of the instability. Using the model proposed by Yerger et al.~\cite{yerger_2025}, the effective parallel heat conductivity is estimated as $\kappa_{\rm Y} \approx 0.13 \kappa_{\rm S}$, while the Ryutov model~\cite{ryutov_1999} with $\ell_{\rm B} \approx \rho_e$ (to match the predicted scale of unstable WHFI modes) yields $\kappa_{\rm R} \approx 0.11 \kappa_{\rm S}$. In addition, strong electron magnetization ($\lambda_e/\rho_e \approx 11$) results in cross-field suppression, $\kappa_{\perp} \approx 0.019 \kappa_{\rm S}$~\cite{braginskii_1965}. These reductions correspond to an effective conductivity of $\kappa_{\text{eff}}/\kappa_{\text{S}} = [(1/3)\kappa_{\|}+(2/3)\kappa_{\perp}]/\kappa_{\text{S}} \approx 0.05$–$0.07$ for the Ryutov and Yerger models, respectively. These estimates are consistent with the suppression of Spitzer-H\"arm conductivity required by our analysis of the thermal energy from the synthetic FLASH data to reproduce the experimental observations. 

To investigate further the sensitivity of the temperature evolution to the plasma thermal conductivity, we performed a second FLASH simulation using an alternative conduction model. In the shocked plasma, we assume that $\kappa_{\rm eff} \approx \kappa_{\rm S} \exp{[-(t-t_0)/\Delta t_{\rm tr}}]+\kappa_{\rm R} \{1-\exp{[-(t-t_0)/\Delta t_{\rm tr}]}\}$, providing a simple model for the transition from classical to microinstability-mediated transport. To match with our experimental observations of when stochastic magnetic fluctuations develop, we take $\Delta t_{\rm tr} = 450~\text{ps}$ and $t_0 = 1.8~\text{ns}$. This delayed-Ryutov model reproduces the experimental temperature evolution more closely (Fig.~\ref{fig:temperature_evolution_and_xRay_spectrum}c, maroon line). Before $\sim2.0~\text{ns}$, transport is governed by the comparatively large Spitzer-H\"arm conductivity, leading to cooling. At later times, Ryutov-like transport dominates, and interactions between the ablation plume and the expanding shock front heat the upper region of the shocked plasma. This produces a hot layer that cannot efficiently equilibrate with the surrounding plasma, resulting in the temperature increase observed at $\sim2.5~\text{ns}$. Although the anomalous conductivity model used in these FLASH simulations is empirically motivated -- a first-principles equivalent would require a self-consistent model of anisotropic heat conduction coupled to whistler-instability evolution~\cite{drake_2021} -- the results confirm that the plasma temperature evolution in our experiment is sensitive to the thermal conductivity, and that a plausible model of whistler-regulated heat conduction reproduces the observations more closely than classical transport. See supplementary information for analysis of synthetic FLASH data simulated with Delayed-Ryutov thermal conduction.

Our results link kinetic microinstabilities to suppressed thermal conduction in a weakly collisional, high-$\beta$ plasma, with important implications for modeling heat conduction in astrophysical, laser-produced, and ICF plasmas. These experiments provide a proof of principle that a process previously investigated primarily by analytical calculations and simulations can also be studied directly in the laboratory. Our measurements therefore provide a benchmark for models of microinstability-mediated thermal transport. Incorporating such models into our FLASH simulations of the experiment would enable quantitative comparison of the various anomalous conductivity models that have been proposed. Recent theoretical and numerical studies of weakly collisional plasmas have shown that anomalous thermal conductivity models can qualitatively alter macroscopic dynamics such as thermal instability~\cite{choudhury_2026b}, underlying the importance of testing such models.  
More directly, our results may have implications for efforts to improve the performance of ICF capsules through externally imposed magnetic fields~\cite{moody_2022}. Current models generally assume that magnetization suppresses cross-field conductive losses while leaving transport along field lines close to the classical Spitzer-H\"arm value, limiting the degree to which thermal conduction can be suppressed~\cite{walsh_2022}.
Yet the hotspots of burning, magnetised ICF capsules are well described as weakly collisional ($\lambda_e/L_{\rm T} \sim 10^{-3}$-$10^{-2}$) and high-$\beta$ ($\beta_e \sim 10^{3}$-$10^{5}$) plasmas, suggesting that heat-flux-driven microinstabilities could also suppress parallel conductivity. Although this possibility requires further investigation under ICF-relevant conditions, such suppression could provide an additional benefit of magnetization and motivate new optimization pathways for future ICF platform designs~\cite{walsh_2025}.  


\section*{Methods}

\subsection*{Magnetic Reynolds number calculation}

The earliest observation of shocked plasma during the campaign was at $1.4~\text{ns}$ after the start of the long-pulse (LP) drive. Given the symmetry of the blow-off expansion, the plasma travelled $L_{\text{f}}\approx 0.13~\text{cm}$, corresponding to the distance between the centre of the drive spot and the top of the shock foil. Assuming constant velocity of the blow-off resulted in the flow velocity quoted in the main text. The magnetic resistivity was estimated using $T_{e}\approx 1.5~\text{keV}$, inferred from the Ly-$\alpha$/He-$\beta$ line ratio in the $0.6~\text{ns}$ spectral data shown in Extended Data Fig. \ref{fig:spectral_line_identification}a, the earliest measured blow-off plasma emission.

\subsection*{Magnetic field reconstructions}

\begin{figure*}
    \centering
    \includegraphics[width=1\linewidth]{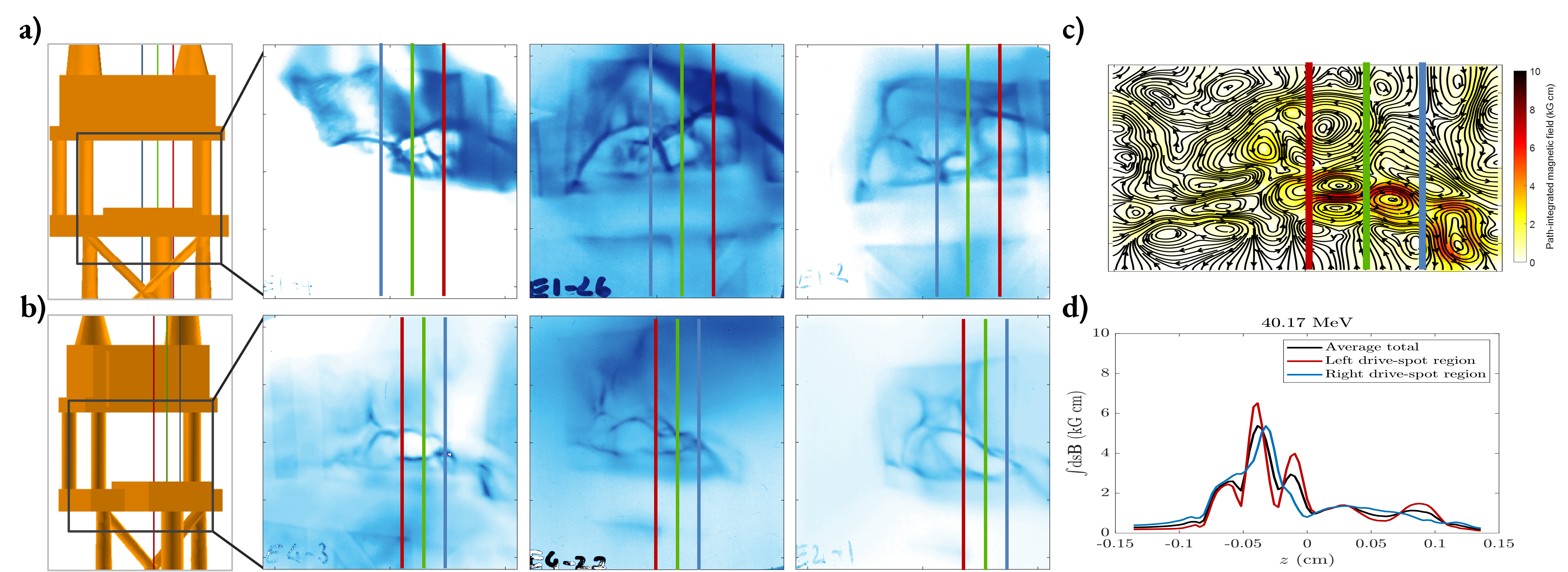}
    \caption{\textbf{PRAD shock foil orientation.} (\textbf{a}) Target visualisation from the SP2 LoS , overlaid with lines corresponding to the LP drive-beam centres (red and blue), and  the green showing the centre of the target. The images to the right are normalised proton fluence images from the same LoS over time. These have similarly been overlaid with the coloured lines to orientate the viewer on the relative position of the signal, and hence, plasma within each image. (\textbf{b}) Same as (a) but for the SP1 LoS. Each fluence image is plotted on a unique colour bar to best show the observed structures. (\textbf{c}) Reconstructed, LoS integrated, magnetic field using the $2.2$ ns, $40.17$ MeV proton image, overlaid with the position lines discussed in (a). (\textbf{d}) Lineouts of the reconstructed, LoS integrated, magnetic field from (c). The red, and blue lines showing the data averaged from a drive spot centre line to the target centre, and then the resultant average of the both indicated by the black line.}
    \label{fig:prad_foil_orientation}
\end{figure*}

The proton fluence images are used to reconstruct the line-of-sight (LoS)-integrated magnetic-field strength and morphology using the method described in ~\cite{bott_2017}. Extended Data Fig.~\ref{fig:prad_foil_orientation}a,b show target visualisations from the perspective of the RCF stack used to record the proton fluence generated by each short-pulse (SP) drive. The visualisations indicate the lines passing through the centres of the LP drive-beam spots and foil centre, which define the regions of interest for the magnetic-field reconstructions.

To align the target geometry with the proton fluence data, the target shadows, most clearly visible in the lower-energy proton images, were aligned with the corresponding target visualisations. Extended Data Fig.~\ref{fig:prad_foil_orientation}c shows the LoS-integrated magnetic-field reconstruction for the $2.2~\text{ns}$ SP2 fluence data and the indicated lines of interest. The lineouts in Extended Data Fig.~\ref{fig:prad_foil_orientation}d were extracted from the LoS-integrated magnetic field and averaged symmetrically about the corresponding centre line.



\subsection*{X-ray self-emission ratio curve use, and detector response}
\begin{figure*}[h]
    \centering
    \includegraphics[width=1\linewidth]{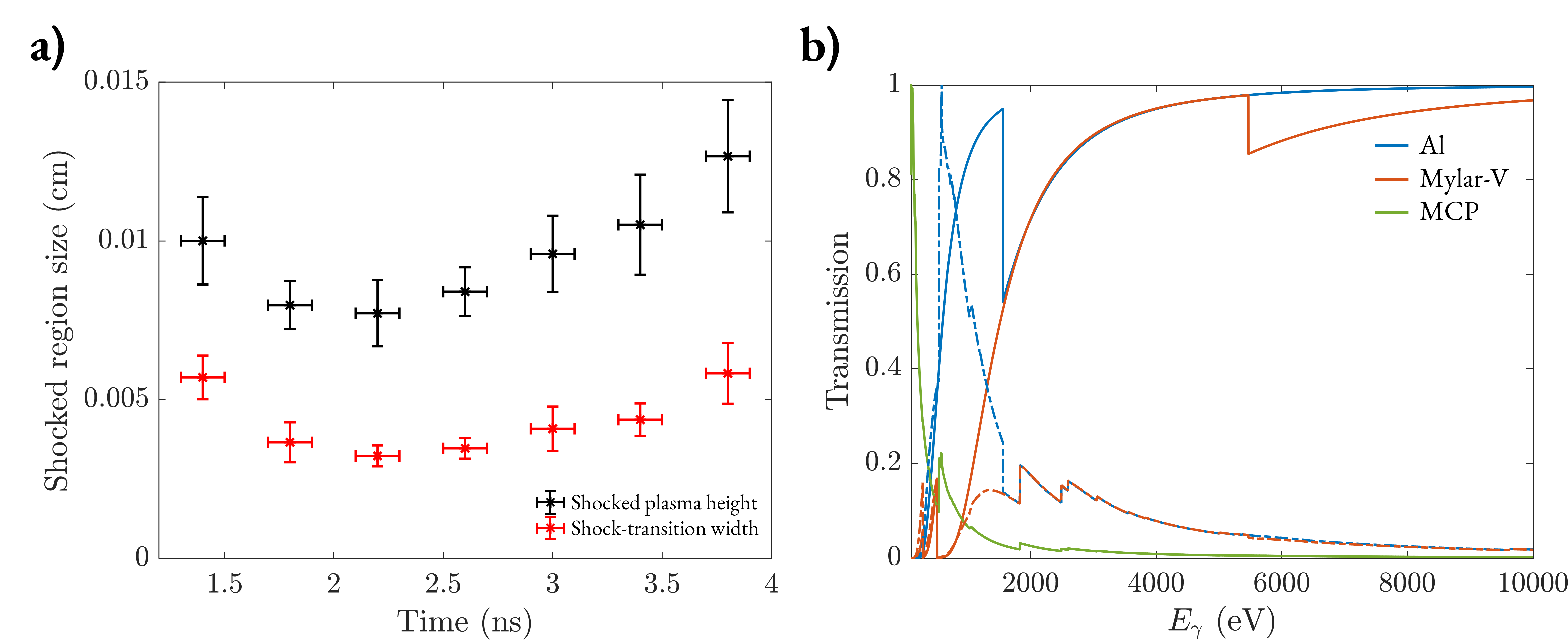}
    \caption{\textbf{X-ray self emission dimensions with detector filter transmission and response.} (\textbf{a}) Temporal evolution of the averaged shocked plasma height, and shock-transition width in black, and red respectively. The vertical error bars correspond to one standard deviation from the mean taken between each filter and repeated shots. The horizontal error bars correspond to the $200~\text{ps}$ intra-strip sweep the detector used when taking the self-emission measurements. (\textbf{b}) Transmission of the Al and Mylar + V filters as a function of photon energy indicated by the solid blue and orange lines, respectively. The green solid line indicates the MCP response function. The blue and orange dashed lines show the transmission of Al and Mylar + V filters, respectively, convolved with the MCP response.}
    \label{fig:gxd_transmission}
\end{figure*}

Each filter-ratio curve is calculated at fixed density, requiring the plasma density to be constant along the diagnostic line of sight (LoS)\cite{poole_2026}. Although this assumption is not made for the produced plasma, density fluctuations along the diagnostic LoS are unlikely to significantly affect the analysis. Fig.~\ref{fig:gxd_temp_map_data}c shows ratio--temperature curves calculated at fixed densities between $0.9$ and $3.0\times10^{20}~\text{cm}^{-3}$. At lower temperatures ($T_{e}\sim500~\text{eV}$), the curve profile is sensitive to density, but this dependence becomes negligible for $T_{e}\gtrsim500~\text{eV}$. Thus, reasonable density fluctuations along the diagnostic LoS do not significantly affect the interpretation of the data. The framing-camera response and filter transmissions are shown in Extended Data Fig.~\ref{fig:gxd_transmission}b.

\subsection*{X-ray self emission signal flat-fielding}\label{sec:flat_fielding}
\begin{figure*}
    \centering
    \includegraphics[width=0.8\linewidth]{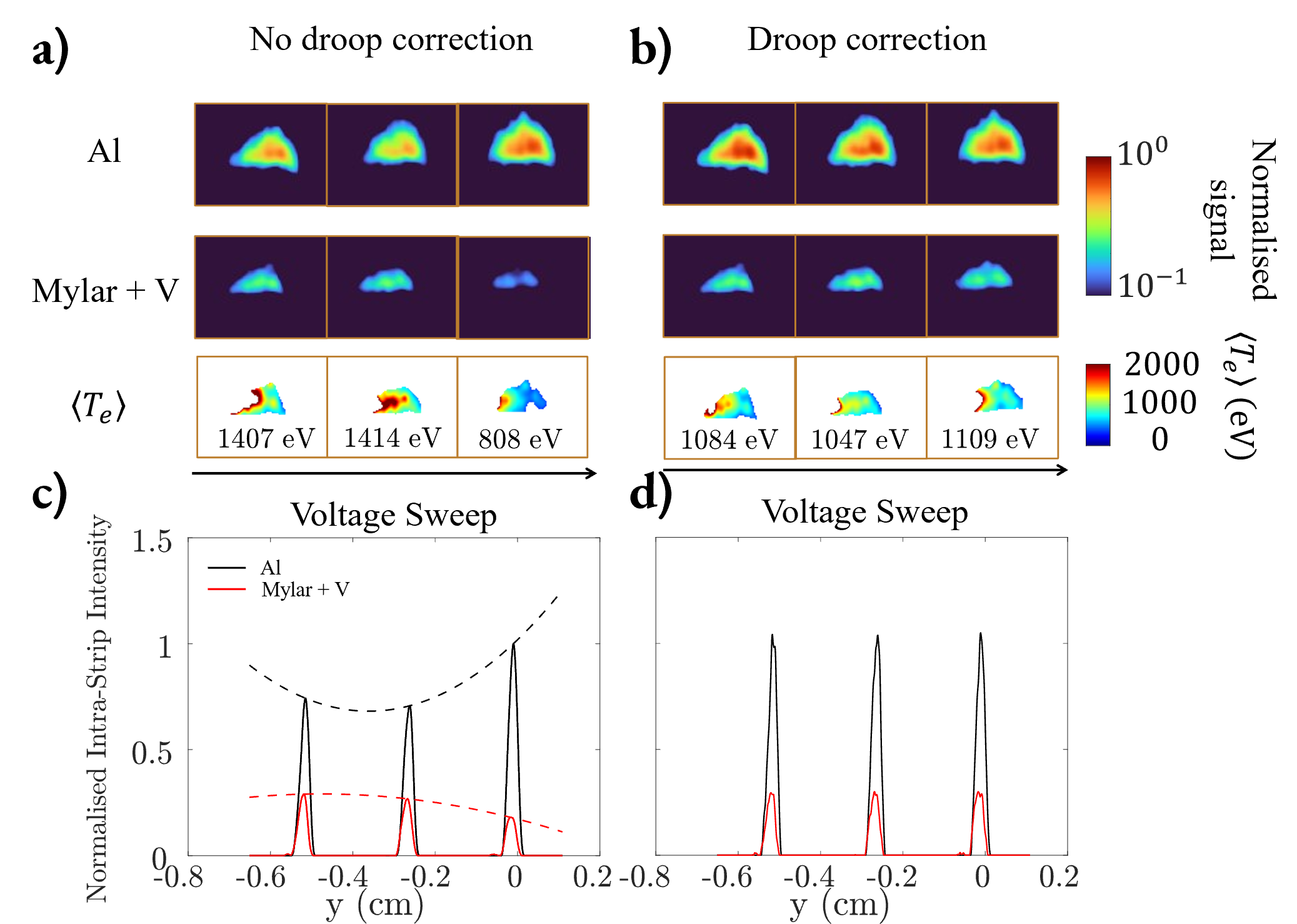}
    \caption{\textbf{Flat-fielding of GXD data.} (\textbf{a}) Self emission x-ray data evolution filtered by 0.544 um Al (top) and 0.9 um Mylar + 0.416 um V (centre) captured along one strip of the MCP array, with the voltage sweep travelling from left to right. Filter ratio inferred temperature maps are shown (bottom) corresponding to the x-ray data in the top and middle columns. (\textbf{b}) Same as (a) but for flat-field corrected data. (\textbf{c})  Lineouts of intensity, integrated in the vertical axis, for the emission data shown in (a). The black and red solid lines correspond to emission from the Al, and Mylar + V filters, respectively. The dashed lines are droop profile inferred from the peak emission, intra-strip, for each filter. (\textbf{d}) Same as (c), but for emission data shown in (b).}
    \label{fig:flat_fielding}
\end{figure*}

The voltage sweep used to obtain the intra-strip framing-camera images can, in some cases, lead to signal droop along the sweep direction~\cite{valenzuela-villaseca_2024}, affecting interpretation of the measured signals. Extended Data Fig.~\ref{fig:flat_fielding}a shows x-ray self-emission data measured through the gated x-ray detector (GXD) for each filter, together with the corresponding temperature maps generated from their ratio. Each strip contains three images, so the signal can vary along a given strip. Since the sweep time is ${\sim}200~\text{ps}$, shorter than any characteristic dynamical timescale in the plasma, significant signal variation during a single imaging period is not expected. However, a non-negligible difference in signal intensity is observed between images in both filters, leading to significant variations in the inferred average temperature, from $808$ to $1414~\text{eV}$ for this shot.

Integrating the signal along the vertical axis and plotting the intensity across the horizontal axis, Extended Data Fig.~\ref{fig:flat_fielding}c shows the intra-strip intensity variation for each filter. The line connecting the intensity peaks defines the droop profile; dividing the signal by this profile corrects the droop across the strip. The corrected signals, temperature maps, and integrated lineouts are shown in Extended Data Fig.~\ref{fig:flat_fielding}b,d. The corrected image intensities are substantially more consistent, with the corresponding lineouts exhibiting nearly constant peak heights. The temperature maps are similarly consistent, with inferred average temperatures ranging from $1047$ to $1109~\text{eV}$.

\subsection*{Plasma emissivity calibration for density estimates}

The plasma emissivity depends on both the temperature and density of the emitting region. By $2.2~\text{ns}$ after the start of the LP drive, the ratio of peak ablation-plume to shocked-plasma emission inferred from the synthetic SPECT3D post-processed FLASH data is significantly larger than that observed experimentally. As the emissivity depends only weakly on temperature over the relevant parameter space, this discrepancy is expected to arise primarily from differences in density, implying that the experimentally generated plasma is denser than predicted by the FLASH simulations.

\begin{figure*}
    \centering
    \includegraphics[width=1\linewidth]{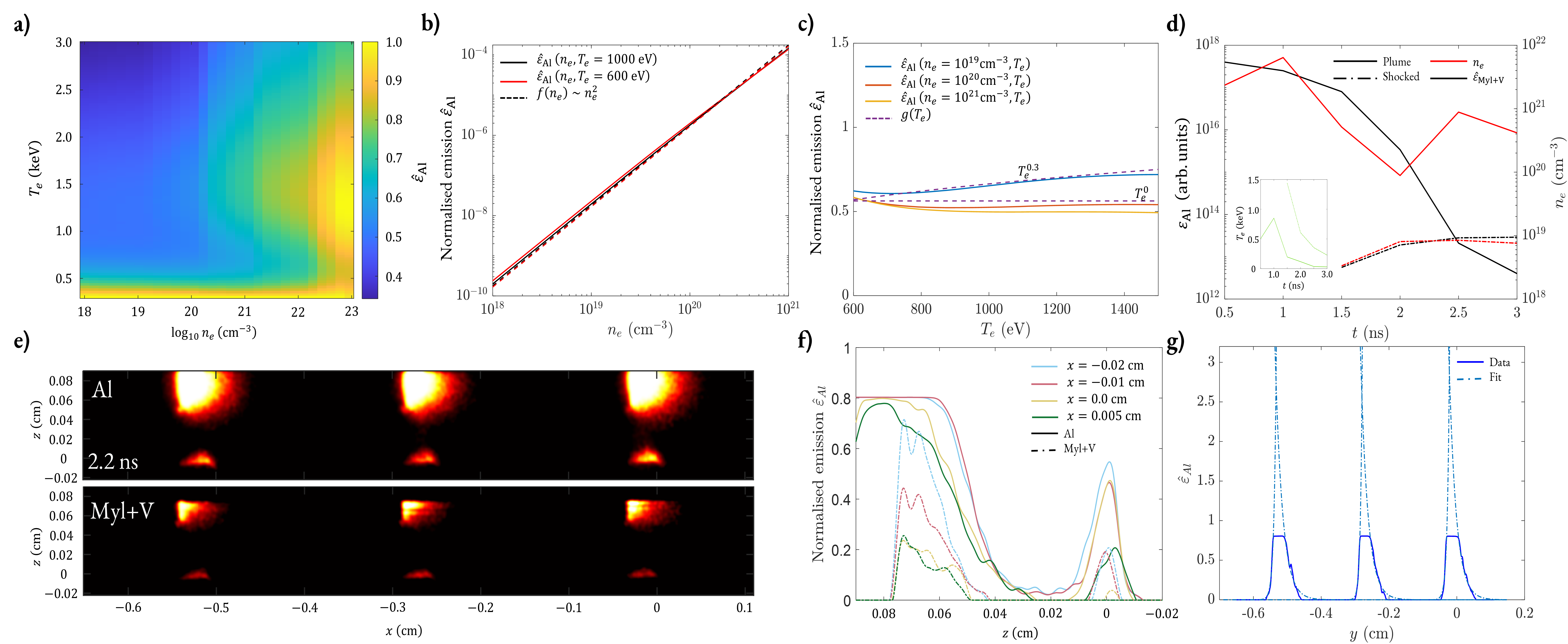}
    \caption{\textbf{Al emissivity modelling.} (\textbf{a}) Normalised emissivity map over temperature and density space, filtered through $0.544~\mu$m Al, and detector response function. At each density, the temperature dependence has been normalised to its maximum value. (\textbf{b}) Normalised emissivity vs. density at fixed $T_{e}=1000~\text{eV}$, and $T_{e}=600~\text{eV}$ shown by the black and red solid lines, respectively. The dashed lines indicate quadratic density fits for each temperature lineout. (\textbf{c}) Normalised emissivity vs. temperature at fixed $n_{e}=10^{19}$-$10^{21}~\text{cm}^{-3}$, indicated by the blue, orange, and yellow solid lines, respectively. The dashed lines indicate approximate temperature scalings for each fixed density. (\textbf{d}) Synthetic FLASH ablation plume, and shocked plasma emissivity (black) through the Al filter $\hat{\varepsilon}_{\text{Al}}$, and $n_{e}$ (red) over time, with the solid lines indicating parameters from the ablation plume, and dashed from the shocked plasma. The inset shows the corresponding evolution of $T_{e}$. (\textbf{e}) Filtered self-emission GXD data at $2.2~\text{ns}$ from the start of the LP drive. The top panel shows the Al, and the bottom the Mylar+V filtered data. (\textbf{f}) Emission line-outs from the Al, and Mylar+V filtered data, shown by the solid and dashed lines, respectively, for a series of positions along $x$. (\textbf{g}) Line-outs of the Al filtered emission data, taken at the $z$ position corresponding to the peak emission, shown by the solid navy line. The blue dash-dotted lines show exponential fits to either side of each saturated emission peak.}
    \label{fig:emissivity_density_callibration}
\end{figure*}

Extended Data Fig. \ref{fig:emissivity_density_callibration}a shows the normalised Al-filtered emissivity, $\hat{\varepsilon}_{\text{Al}}$, as a function of $n_{e}$ and $T_{e}$. Motivated by these trends, the plume emissivity may be expressed relative to that of the shocked plasma as
\begin{equation}
    \varepsilon_{\text{plume}}\propto f(n_{e})g(T_{e})\varepsilon_{\text{shock}},
\end{equation}
where $\varepsilon_{\text{plume}}$ and $\varepsilon_{\text{shock}}$ are the emission from the blow-off plume and shocked plasma, respectively, and $f(n_{e})$ and $g(T_{e})$ describe the respective density and temperature dependencies of the shocked-plasma emission. This relation allows the plume-to-shocked-plasma emission ratio inferred from the FLASH simulations to be compared directly with experiment:
\begin{equation}
    \left(\frac{\varepsilon_{\text{plume}}}{\varepsilon_{\text{shock}}}\right)_{\text{FLASH}} = \frac{\left[ f(n_{e})g(T_{e}) \right]_{\text{exp}} }{\left[ f(n_{e})g(T_{e}) \right]_{\text{FLASH}}}\left(\frac{\varepsilon_{\text{plume}}}{\varepsilon_{\text{shock}}}\right)_{\text{exp}},
    \label{eq:emission_ratio_dependence}
\end{equation}
where the subscripts FLASH and exp refer to the synthetic and experimentally inferred parameters, respectively.

Extended Data Fig. \ref{fig:emissivity_density_callibration}b,c show emissivity lineouts at fixed temperature and density, respectively, illustrating its density and temperature dependence over ranges representative of the inferred plasma conditions. The emissivity exhibits an approximately quadratic dependence on density and is effectively independent of temperature for $n_{e}\gtrsim10^{20}~\text{cm}^{-3}$, while at lower densities it has a weak temperature dependence, $g(T_{e})\sim T_{e}^{0.3}$ at $n_{e}=10^{19}~\text{cm}^{-3}$. Eq.~(\ref{eq:emission_ratio_dependence}) therefore reduces to
\begin{equation}
    n_{e,\text{exp}}=n_{e,\text{FLASH}}\sqrt{\frac{\left(\varepsilon_{\text{plume}} / \varepsilon_{\text{shock}}\right)_{\text{FLASH}}}{ \left(\varepsilon_{\text{plume}}  / \varepsilon_{\text{shock}}\right)_{\text{exp}}}},
    \label{eq:emission_calibration_density}
\end{equation}
providing an estimate of the shocked-plasma density.

Extended Data Fig. \ref{fig:emissivity_density_callibration}d shows the evolution of $\hat{\varepsilon}{\text{Al}}$, $n_{e}$, and $T_{e}$ for the ablation plume and shocked plasma in the synthetic FLASH data. Plume parameters are sampled from the region spatially coincident with the peak emission, while shocked-plasma parameters are averaged over regions exceeding $50\%$ of the local maximum emission within a spatial window centred on the shocked plasma, as described in the main text (see Fig.~\ref{fig:gxd_temp_map_data}). These results provide the calibration for Eq.~(\ref{eq:emission_calibration_density}). Extended Data Fig. \ref{fig:emissivity_density_callibration}e shows GXD data from the Al- and Mylar-V-filtered images at $2.2~\text{ns}$ after the start of the LP drive. Each strip has been flat-field corrected as described in Section~\ref{sec:flat_fielding}. In both filtered datasets, the shocked plasma exhibits lower emissivity than the ablation plume, consistent within the emissivity model with a higher shocked-plasma density. Owing to the strip orientation, the Mylar-V image is truncated at the ablation plume, so the peak plume emission cannot be measured reliably or used directly for density calibration.

Extended Data Fig. \ref{fig:emissivity_density_callibration}f shows lineouts along $z$ at different $x$ positions for the rightmost GXD emission image for each filter. The Mylar-V-filtered plume emission has a smaller vertical extent than the Al-filtered emission, with a sharper drop-off due to the imaging-strip edge. As the shocked-plasma peak emission overlaps spatially between the filters, the plume emission is also expected to overlap. Furthermore, owing to the vertical LP-drive symmetry, the $z$ position of the peak plume emission should remain constant along $x$. At $x=0.005~\text{cm}$, the Al-filtered plume peaks at $z\approx0.08~\text{cm}$, whereas the Mylar-V-filtered plume peaks at $z<0.08~\text{cm}$. Thus, Eq.~(\ref{eq:emission_calibration_density}) is not applicable to the Mylar-V-filtered data.

The Al-filtered plume emission also saturates the GXD display and requires correction before density estimation. Extended Data Fig. \ref{fig:emissivity_density_callibration}g shows lineouts along $x$ at $z=0.08~\text{cm}$ from the Al-filtered emission in Extended Data Fig. \ref{fig:emissivity_density_callibration}e. In all intra-strip images, the ablation-plume emission saturates at approximately 0.8 in normalised units. Exponential fits to the emission profiles on either side of the saturation plateau are overlaid, and their intersections provide estimates of the unsaturated peak emission. The ablation-plume-to-shocked-plasma brightness ratio is then evaluated using the FLASH-inferred quantities between $1.5$ and $2.0~\text{ns}$. The quoted density is the mean of the resulting estimates over this interval, with the uncertainty given by their spread.

\subsection*{Estimate of Coulomb mean free path of ions in the ablation plume plasma through the shocked plasma}

For ablation plume ions penetrating the shocked plasma, Coulomb interactions with all other ion species and electrons can, in principle, slow them down. Because of its atomic composition, the momentum of the ablation plume plasma is predominantly contributed to by its carbon ions, and so their slowing is most pertinent to shock formation. We therefore compute the slowing-down rates $\nu_{\rm s}^{\mathrm{C}|\mathrm{C}}$, $\nu_{\rm s}^{\mathrm{C}|\mathrm{H}}$, and $\nu_{\rm s}^{\mathrm{C}|e}$ of carbons with other carbons, hydrogens, chlorines, and electrons respectively. Because the ablation plume is moderately supersonic, the carbon ions are fast with respect to ions in the shocked plasma, but slow with respect to the electrons~\cite{Beresnyak_2023}. It follows that
\begin{eqnarray}
\nu_{\rm s}^{\mathrm{C}|\mathrm{C}} & \approx & 1.8 \times 10^{-7} M_{\rm C}^{1/2} \epsilon_{\rm C}^{-3/2} n_{\rm C} Z_{\rm C}^4 \log \Lambda_{\rm CC} \, , \\
 \nu_{\rm s}^{\mathrm{C}|\mathrm{H}} & \approx & 9.0 \times 10^{-8} M_{\rm C}^{-1/2} \epsilon_{\rm C}^{-3/2} n_{\rm H} \nonumber \\
 && \qquad \times (1+M_{\rm C}/M_{\rm H}) Z_{\rm C}^2 Z_{\rm H}^2 \log \Lambda_{\rm CH} \, ,  \\
 \nu_{\rm s}^{\mathrm{C}|\mathrm{Cl}} & \approx & 9.0 \times 10^{-8} M_{\rm C}^{-1/2} \epsilon_{\rm C}^{-3/2} n_{\rm Cl} \nonumber \\
 && \qquad \times (1+M_{\rm C}/M_{\rm Cl}) Z_{\rm C}^2 Z_{\rm Cl}^2 \log \Lambda_{\rm CCl} \, ,  \\
\nu_{\rm s}^{\mathrm{C}|e} & \approx & 1.6 \times 10^{-9} M_{\rm C}^{-1} T_e^{-3/2} n_e Z_{\rm C}^2 \log \Lambda_{\mathrm{C}e} \, . 
\end{eqnarray}
Here, $M_{\rm C} = 12$, $M_{\rm H} = 1$, and $M_{\rm Cl} = 1$ are the relative atomic masses of carbon, hydrogen, and chlorine; $Z_{\rm C} = 6$, $Z_{\rm H} = 1$, and   $Z_{\rm Cl} = 15$ are the estimated charge of carbon, hydrogen, and chlorine in the shocked plasma; $n_{\rm s’}$ the number density of species $s$ (in $\mathrm{cm}^{-3}$); $\epsilon_{\rm C} = M_{\rm C} m_p u_{\rm plume}^2/2$ is the energy of the ablation plume carbons (in eV), for $m_p$ the proton mass; and $\log{\Lambda_{ss'}}$ the Coulomb logarithm of species $s$ interacting with species $s'$. 

We then evaluate these slowing-down rates for parameters characteristic of our experiment at $t = 2.2$ ns after the start of the long-pulse laser drive. We estimate the ablation plume velocity by $u_{\rm plume} \approx \Delta L/t\approx 5.4 \times 10^{7}$ cm/s, where $\Delta L \approx L_{\rm f} - L_{\rm int} \approx 0.12$ cm is the distance travelled by ablation plume plasma from the long-pulse laser spots to the top of the shocked plasma. The slowing-down rates are then
\begin{eqnarray}
\nu_{\rm s}^{\mathrm{C}|\mathrm{C}} & \approx & 2.4 \times 10^{8} \left[\frac{n_e (\mathrm{cm}^{-3})}{10^{20} \; \mathrm{cm}^{-3}}\right]  \nonumber \\&& \quad \times  \left[\frac{u_{\rm plume}(\mathrm{cm/s})}{5.4 \times 10^{7} \, \mathrm{cm/s}}\right]^{-3} \left[\frac{\log \Lambda_{\rm CC}}{6}\right] \, \mathrm{s}^{-1}, \\
 \nu_{\rm s}^{\mathrm{C}|\mathrm{H}} & \approx & 5.2 \times 10^{8} \left[\frac{n_e (\mathrm{cm}^{-3})}{10^{20} \; \mathrm{cm}^{-3}}\right] \nonumber \\&& \quad \times \left[\frac{u_{\rm plume}(\mathrm{cm/s})}{5.4 \times 10^{7} \, \mathrm{cm/s}}\right]^{-3} \left[\frac{\log \Lambda_{\rm CH}}{8}\right] \, \mathrm{s}^{-1}, \\
 \nu_{\rm s}^{\mathrm{C}|\mathrm{Cl}} & \approx & 1.0 \times 10^{9} \left[\frac{n_e (\mathrm{cm}^{-3})}{10^{20} \; \mathrm{cm}^{-3}}\right] \nonumber \\&& \quad \times  \left[\frac{u_{\rm plume}(\mathrm{cm/s})}{5.4 \times 10^{7} \, \mathrm{cm/s}}\right]^{-3} \left[\frac{\log \Lambda_{\rm CCl}}{5}\right] \, \mathrm{s}^{-1}, \\
\nu_{\rm s}^{\mathrm{C}|e} & \approx & 1.3 \times 10^{8} \left[\frac{n_e (\mathrm{cm}^{-3})}{10^{20} \; \mathrm{cm}^{-3}}\right] \nonumber \\&& \quad \times  \left[\frac{T_{e}(\mathrm{eV})}{1.1 \, \mathrm{keV}}\right]^{-3/2} \left[\frac{\log \Lambda_{\mathrm{C}e}}{8}\right] \, \mathrm{s}^{-1} .
\end{eqnarray}
Because the slowing down of the ablation plume's carbon ions by electrons is much less efficient than by the ions, the rate of change of their velocity $v_{\rm C}$ with distance $x$ is approximately given by
\begin{equation}
\frac{\mathrm{d} v_{\rm C}}{\mathrm{d}x} \approx -\left[\nu_{\rm s}^{\mathrm{C}|\mathrm{C}}(v_{\mathrm{C}0})+\nu_{\rm s}^{\mathrm{C}|\mathrm{H}}(v_{\mathrm{C}0})+\nu_{\rm s}^{\mathrm{C}|\mathrm{Cl}}(v_{\mathrm{C}0})\right] \frac{v_{\mathrm{C}0}^3}{v_{\rm C}^3} , 
\end{equation}
where $v_{\mathrm{C}0} = u_{\rm plume}$ is the initial speed of the carbon ions. It follows that the mean free path $\lambda_{\mathrm{C,plume}}$ of the carbons is given by 
\begin{equation}
\lambda_{\mathrm{C,plume}} = \frac{u_{\rm plume}}{4 \left[\nu_{\rm s}^{\mathrm{C}|\mathrm{C}}(u_{\rm plume})+\nu_{\rm s}^{\mathrm{C}|\mathrm{H}}(u_{\rm plume})+\nu_{\rm s}^{\mathrm{C}|\mathrm{Cl}}(u_{\rm plume})\right]} .    
\end{equation}
Evaluating this for the ablation plume parameters gives $\lambda_{\mathrm{C,plume}} \approx 4 \times 10^{-3} \; \mathrm{cm}$, consistent with the shock-transition width reported in the main text.

\subsection*{Spectral line identification}
\begin{figure*}
    \centering
    \includegraphics[width=1\linewidth]{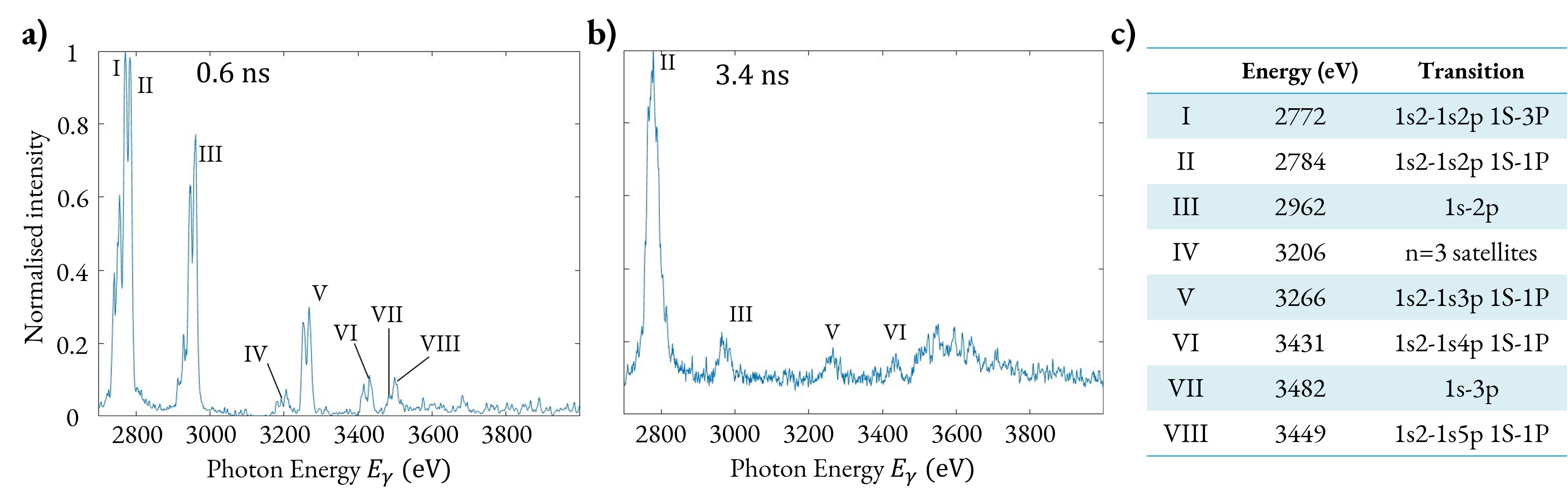}
    \caption{\textbf{Spectral line identification.} (\textbf{a}) Self-emission x-ray spectrum taken from the ablation plasma plume at $0.6$ ns from the start of the LP drive. (\textbf{b}) Self-emission x-ray spectrum taken from the shocked plasma at $3.4$ ns from the start of the LP drive. (\textbf{c}) Photon energy corresponding to the labelled lines, along with their corresponding energy level transitions.}
    \label{fig:spectral_line_identification}
\end{figure*}

Highly resolved spectral lines are observed in the early-time spectroscopy data, originating from the ablation plume. Extended Data Fig.~\ref{fig:spectral_line_identification}a shows emission data acquired at $0.6~\text{ns}$ after the start of the LP drive. To improve the signal-to-noise ratio, the spectra were spatially integrated, with the spectral images used to determine the relative emission position. At later times, reduced signal intensity resulted in loss of spatial resolution, so removing the slit increased the measured signal. Consequently, all data from $2.2~\text{ns}$ onwards used for temperature inference are spatially integrated. Extended Data Fig.~\ref{fig:spectral_line_identification}b shows spectral lines measured at $3.0~\text{ns}$.

Extended Data Fig.~\ref{fig:spectral_line_identification}c lists the photon energy at the centre of each identified line and the corresponding atomic transition. For the $3.0~\text{ns}$ data, where the spectral lines are broader and less well resolved, the dominant transition is used as the label. For example, the intercombination and resonance He-like Cl lines are individually resolved at $0.6~\text{ns}$ but not at $3.0~\text{ns}$; therefore, the peak is labelled using the resonance-line transition.

The plasma of interest is the shocked plasma rather than the ablation plume, therefore from $2.2~\text{ns}$ emission between the two sources must be distinguished. Using a combination of synthetic FLASH~\cite{fryxell_2000} data post-processed with SPECT3D~\cite{macfarlane_2007} with the spatially resolved plume spectra obtained at early times, we conclude that emission exhibiting clear Ly-$\alpha$ lines from $2.2~\text{ns}$ onwards is most likely produced by the shocked plasma rather than the ablation plume.

\begin{figure*}
    \centering
    \includegraphics[width=1\linewidth]{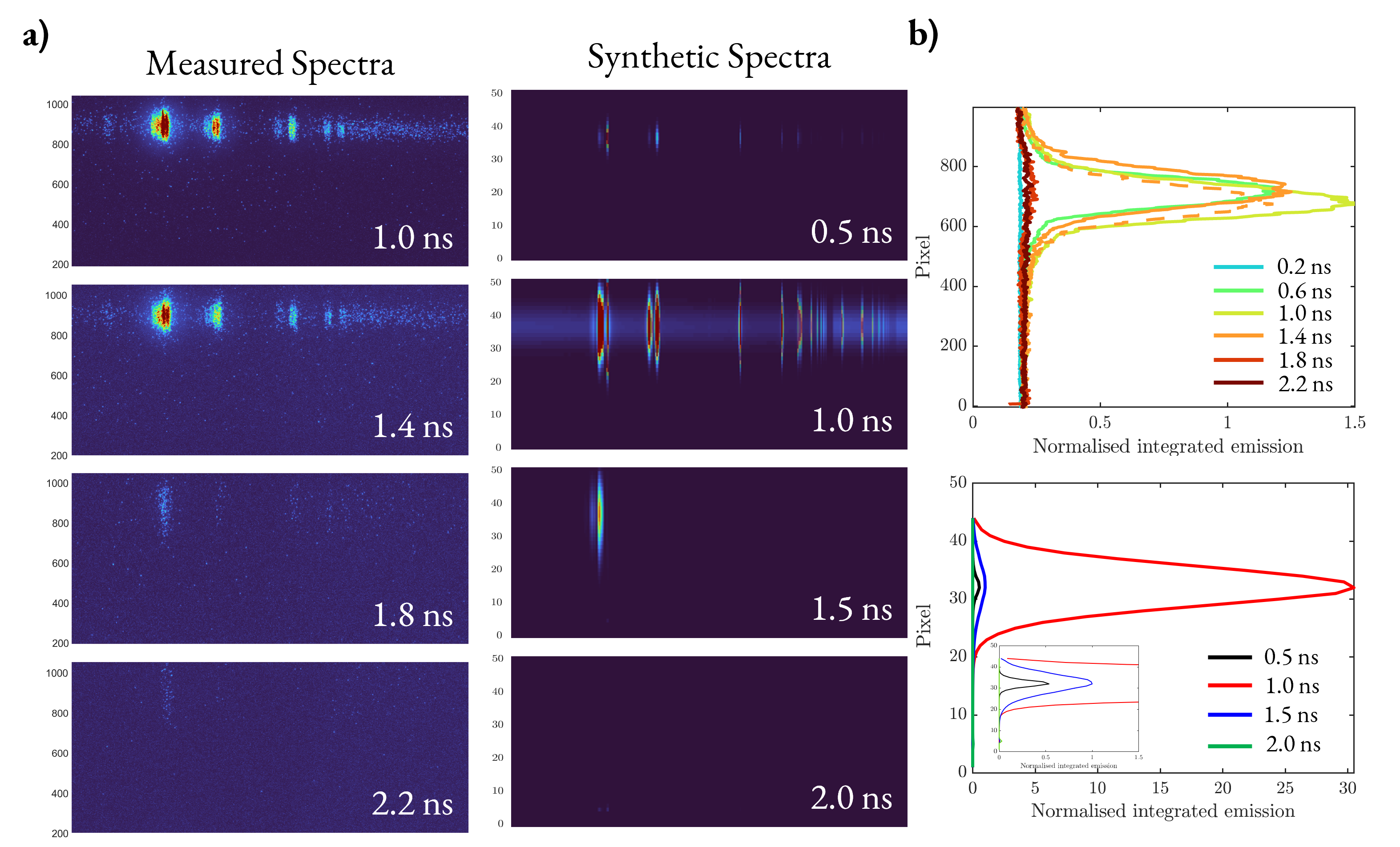}
    \caption{\textbf{Evolution of peak emissivity.} (\textbf{a}) Spatially resolved spectra over time for the experimentally measured data (left), and post-processed synthetic data (right). (\textbf{b}) spectrally integrated lineouts in the vertical axis for the measured spectra (top), and synthetic post-processed data (bottom). The dashed purple line for the integrated measured intensity, correspond to the camera bias-corrected $1.4~\text{ns}$ data. The insert for the integrated synthetic intensity, shows the intensities on the same scale as the measured data.}
    \label{fig:emissive_region_over_time}
\end{figure*}

Extended Data Fig.~\ref{fig:emissive_region_over_time}a shows the measured spectra, spatially resolved along the vertical axis, between $1.0$-$2.2~\text{ns}$ after the start of the LP drive, together with synthetic FLASH spectra post-processed using SPECT3D between $0.5$-$2.0~\text{ns}$ over the same spectral range. The measured $1.0~\text{ns}$ data were acquired using a $-200~\text{V}$ camera bias, compared with the $-250~\text{V}$ bias used for the other spectra shown.

In both the datasets, the early-time peak emission is located at the centre of the drive foil and decreases rapidly after $\sim1.5~\text{ns}$. The spectral images are normalised to the peak emission of the $1.4~\text{ns}$ measured data and the $1.5~\text{ns}$ synthetic data, respectively. Extended Data Fig.~\ref{fig:emissive_region_over_time}b shows the spatially resolved, spectrally integrated intensity profiles for the measured and synthetic data between $0.2$-$2.2~\text{ns}$ and $0.5$-$2.0~\text{ns}$ after the start of the LP drive, respectively. For the measured lineouts, data between $0.2$-$1.4~\text{ns}$ were acquired using a $-200~\text{V}$ bias, while those between $1.4$-$2.2~\text{ns}$ used $-250~\text{V}$. To account for the resulting change in relative brightness, the background levels of the $-250~\text{V}$ data were adjusted such that the normalised heights of the two $1.4~\text{ns}$ peaks were equal. The corrected $-250~\text{V}$, $1.4~\text{ns}$ lineout is shown by the dashed line in Extended Data Fig.~\ref{fig:emissive_region_over_time}b.

The lineouts confirm that the peak brightness originates from the centre of the drive foil at early times and decreases rapidly after $\sim1.5~\text{ns}$. Although the synthetic emission is clearly stronger than that measured experimentally, particularly at $1.0~\text{ns}$, both datasets exhibit a similarly rapid decrease in drive-foil emissivity following the laser drive. At $2.0~\text{ns}$, the synthetic spectra show weak emission in the region corresponding to the shocked plasma, indicating a transition in the dominant emissive region from the drive foil to the shock foil. This transition is not directly observed in the measured data; however, given the weak synthetic signal, the corresponding experimental emission could plausibly have been below the detection threshold.

The spectrometer was fielded with a $25~\mu\text{m}$ beryllium filter and a $5~\text{mm}\times20~\mu\text{m}$ slit, both of which may have attenuated the measured signal, although the lines of interest are optically thin to the former. The slit was subsequently removed, increasing the aperture to a $10~\text{mm}$ diameter collection window and the measured signal at the expense of spatial resolution. The rapid decline of the ablation-plasma emission and the transition in emissivity observed in the synthetic spectra therefore support the assumption that the spatially integrated spectral data acquired from $2.2~\text{ns}$ onwards originate predominantly from the shocked plasma. All measured spectra were subject to the same time integration as in the GXD analysis.

\subsection*{Spectral line fitting}

\begin{figure*}
    \centering
    \includegraphics[width=1\linewidth]{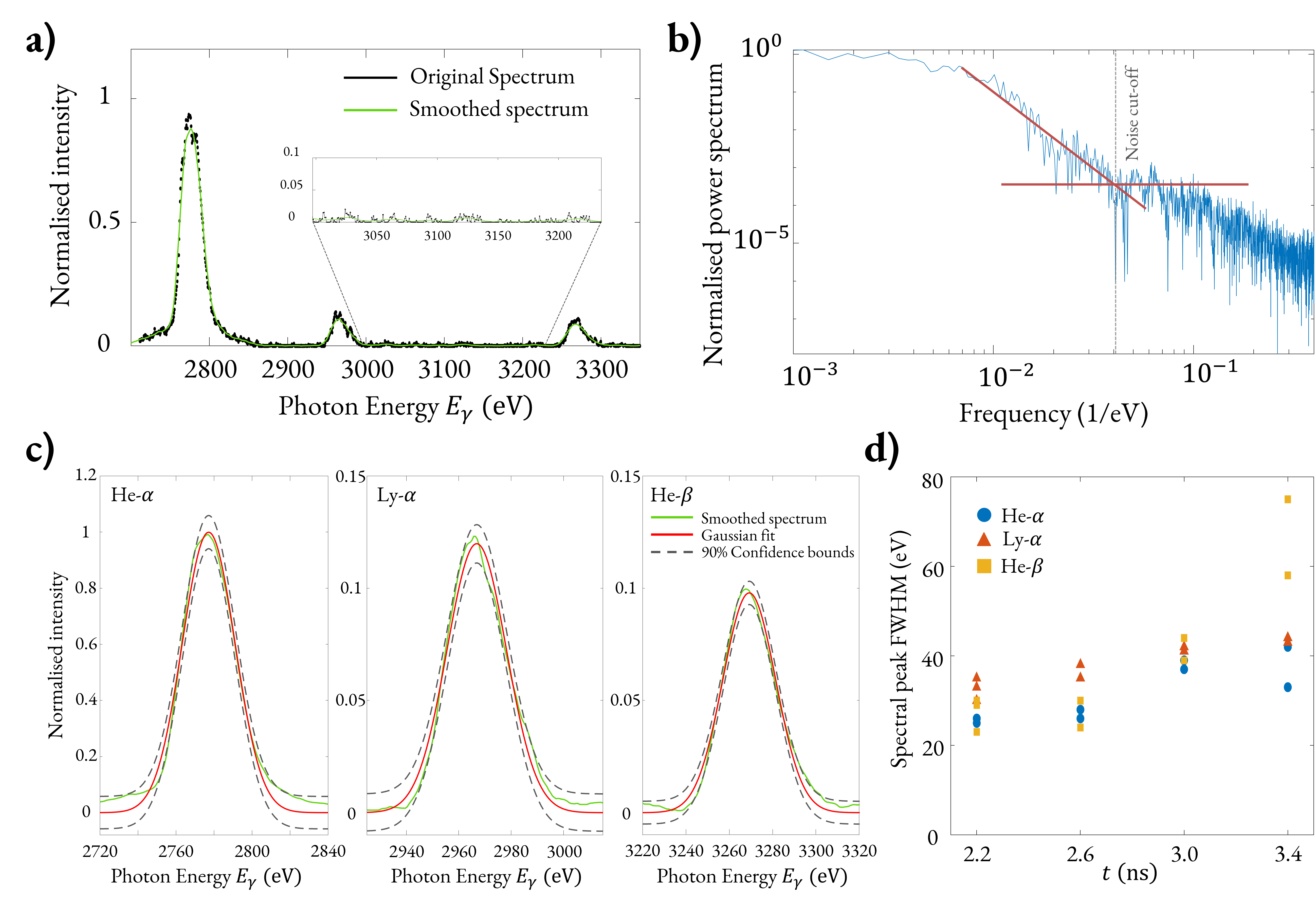}
    \caption{\textbf{Spectrum smoothing and peak fitting.} (\textbf{a}) Self-emission x-ray spectrum taken from the shocked plasma at n ns from the start of the LP drive. The black line shows the raw x-ray spectrum normalised to the He-$\alpha$ peak, and the green line is the noise removed, smoothed spectrum. The insert shows the small scale stochasticity of the noise. (\textbf{b}) Normalised power spectrum of the spectrum shown in (a). The red lines highlight the initial slope of the power spectrum before reaching an initial plateau. Their intersection corresponds to the, noise cut-off, scale over which the smoothing has been performed, indicated by the annotated grey dashed line. (\textbf{c}) He-$\alpha$, Ly-$\alpha$, and He-$\beta$ smoothed spectral peaks from (a) in green, overlaid with Gaussian fits in red and 90\% confidence bounds to the fit shown by the grey dashed lines. All peaks have been normalised to the soothed He-$\alpha$ peak.}
    \label{fig:spectral_widths_and_noise}
\end{figure*}

To infer the relative spectral peak heights, the data are first smoothed. Extended Data Fig.~\ref{fig:spectral_widths_and_noise}a shows a spectrum before and after smoothing, with the inset highlighting small-scale stochastic noise fluctuations. The corresponding power spectrum, shown in Extended Data Fig.~\ref{fig:spectral_widths_and_noise}b, reveals two distinct plateaus. The first corresponds to the detector-resolution noise floor, while the second, annotated in the figure, corresponds to the characteristic scale of the stochastic noise visible in the inset of Extended Data Fig.~\ref{fig:spectral_widths_and_noise}a. The reciprocal of the frequency at the second plateau defines the energy interval over which a moving-average is applied to the spectra.

Gaussian profiles are then fitted to the smoothed spectral peaks using linear least-squares regression. A $90\%$ confidence interval from the fit is used to estimate the uncertainty in the inferred peak-height ratios. Examples of the fits for the He-$\alpha$, Ly-$\alpha$, and He-$\beta$ lines are shown in Extended Data Fig.~\ref{fig:spectral_widths_and_noise}c. Spectral widths are calculated as the full width at half maximum (FWHM) of the fitted Gaussian profiles, with their temporal evolution shown in Extended Data Fig.~\ref{fig:spectral_widths_and_noise}d.

\subsection*{Spectral line ratios and temperature inference}

\begin{figure*}
    \centering
    \includegraphics[width=1\linewidth]{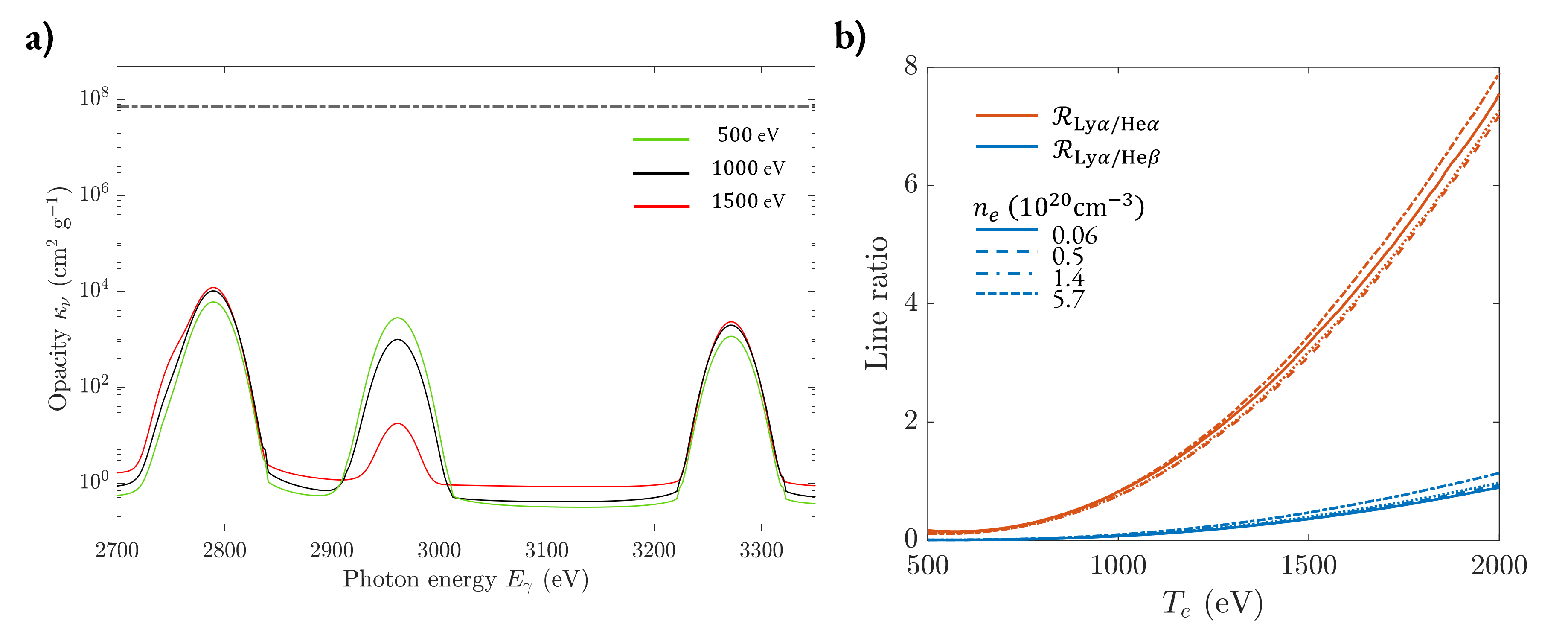}
    \caption{\textbf{Opacity, temperature, and density dependency of emission.} (\textbf{a}) Opacity of He-$\alpha$, Ly-$\alpha$, and He-$\beta$ spectral Cl lines assuming a $30~\text{eV}$ spectral resolution shown between $500$-$1500~\text{eV}$ at constant $n_{e}=1.4\times10^{20}~\text{cm}^{-3}$. The dot-dashed grey line indicates the opacity at which the spectral mean-free path within the plasma is the extent of the shocked region. (\textbf{b}) Line ratio vs. $T_{e}$ for Ly-$\alpha$ to He-$\alpha$, and Ly-$\alpha$ to He-$\beta$ ratios shown in blue, and orange, respectively. Each line ratio-temperature curve is plotted for a range of densities.}
    \label{fig:opacity_and_line_ratios}
\end{figure*}

The opacities of the spectral lines of interest, accounting for their experimentally determined spectral broadening, are shown in Extended Data Fig.~\ref{fig:opacity_and_line_ratios}a. The broadband opacity calculated using PrismSPECT~\cite{golovkin_2025} is shown for temperatures between $500$--$1500~\text{eV}$ at a fixed density of $n_{e}=1.4\times10^{20}~\text{cm}^{-3}$. These opacities provide an estimate of the minimum path length required for significant signal attenuation and hence whether opacity effects must be considered in the line-height ratio analysis.

Using the definition of opacity~\cite{colvin_2013},
\begin{equation}
\rho\kappa_{\nu} = \sum_{j=n_{\text{min}}}^{\infty}n_{j}\sigma_{\nu,j},
\label{eq:opacity_definition}
\end{equation}
where $\rho$ is the plasma mass density, $\kappa_{\nu}$ is the opacity, $n_{j}$ is the population density of bound state $j$, and $\sigma_{\nu,j}$ is the spectral cross-section, and assuming a near-fully ionised plasma dominated by He-like Cl, the He-$\alpha$ transition can be approximated as
\begin{equation}
\rho\kappa_{\nu}\sim n_{i}\sigma_{\nu}\sim \frac{1}{\ell},
\label{eq:opacity_approximation}
\end{equation}
where $\ell$ is the characteristic spectral path length before significant attenuation. For a mixture of ion species, the relative concentration of  species $\alpha$ and $\beta$ is $n_{\alpha}=p_{\alpha}n_{\beta}/p_{\beta}$, where $p_{\alpha}$ is the number fraction of ion species $\alpha$. Combining this relation with the quasineutrality condition,
\begin{equation}
n_{e}=\sum_{\alpha}Z_{\alpha}n_{\alpha},
\end{equation}
where $Z_{\alpha}$ is the ionisation state of ion species $\alpha$, gives the chlorine ion density
\begin{equation}
n_{\text{Cl}} = \frac{p_{\text{Cl}}n_{e}}{\sum_{\alpha} Z_{\alpha}p_{\alpha}}.
\label{eq:chlorine_density}
\end{equation}
together with $\kappa_{\nu}\approx10^{4}~\text{cm}^{2}\text{g}^{-1}$ for the He-$\alpha$ transition at $T_{e}=1000~\text{eV}$ and $n_{e}=1.4\times10^{20}~\text{cm}^{-3}$ (Extended Data Fig.~\ref{fig:opacity_and_line_ratios}a), gives a minimum X-ray path length of $\ell\approx0.86~\text{cm}$ before significant attenuation. This length scale is substantially larger than the extent of the shocked plasma along the spectrometer LoS. The plasma can therefore be considered optically thin for the spectral lines of interest.

The Ly-$\alpha$ to He-$\alpha$ and Ly-$\alpha$ to He-$\beta$ line ratios are used to infer the shocked-plasma temperature. Their temperature dependence, accounting for the experimentally inferred spectral widths, is shown in Extended Data Fig.~\ref{fig:opacity_and_line_ratios}b. The curves were generated by simulating a series of single-cell spectra using PrismSPECT over the relevant temperature and density parameter space and calculating the corresponding line ratios. For both ratios, the temperature dependence is substantially stronger than the density dependence; therefore, a constant plasma density has been assumed. The error bars on the spectroscopy-inferred temperatures quoted in the main text correspond to the range inferred from the minimum and maximum ratios,
\begin{equation}
\mathcal{R}_{\text{min}} =
\frac{I_{\text{Ly-}\alpha,\text{min}}}
{I_{\text{He-}\alpha,\text{max}}},
\end{equation}
and
\begin{equation}
\mathcal{R}_{\text{max}} =
\frac{I_{\text{Ly-}\alpha,\text{max}}}
{I_{\text{He-}\alpha,\text{min}}},
\end{equation}
respectively, using the confidence bounds of the Gaussian fits. For times containing multiple temperature values, their uncertainties were added in quadrature.

\subsection*{Thermal-energy balance analysis}

To infer the theoretical thermal-conductivity suppression from the synthetic FLASH data with flux-limited Spitzer thermal conduction, as described in the main text, an energy balance over the shocked plasma was performed.

The evolution of the electron population within the plasma satisfies the following energy balance~\cite{tzeferacos_2015}:
\begin{equation}
\begin{aligned}
\frac{\text{d}p_{\text{th},e}}{\text{d}t}
&= - \mathbf{\nabla}\cdot\boldsymbol{q}_{e}
- \mathbf{\nabla}\cdot(p_{\text{th},e}\boldsymbol{u}) \\
& - p_{\text{th},e}\mathbf{\nabla}\cdot\boldsymbol{u} + \frac{3k_{B}n_{e}}{2\tau_{ei}}(T_{i} - T_{e})
+ \mathcal{S}_{\text{rad}},
\end{aligned}
\label{eq:thermal_evolution}
\end{equation}
where $p_{\text{th},e} = 3k_{B}n_{e}T_{e}/2$ is the electron thermal pressure, $\tau_{ei}$ is the electron-ion collision time, $T_{i}$ is the ion temperature, $\boldsymbol{q_{e}}$ is the heat flux, $\boldsymbol{u}$ is the plasma flow velocity, and $\mathcal{S_{\text{rad}}}$ is the combined radiative source and sink term.

Integrating over the plasma volume gives
\begin{equation}
\begin{aligned}
\frac{\text{d}}{\text{d}t} \int p_{\text{th},e}~\text{d}V
&= -\int \boldsymbol{q}_{e} \cdot \hat{\boldsymbol{n}}~\text{d}A \\
&- \int p_{\text{th},e}\boldsymbol{u} \cdot \hat{\boldsymbol{n}}~\text{d}A  - \int p_{\text{th},e}\mathbf{\nabla}\cdot\boldsymbol{u}~\text{d}V \\
&+ \int \frac{3k_{B}n_{e}}{2\tau_{ei}}(T_{i} - T_{e})~\text{d}V + \int \mathcal{S}_{\text{rad}}~\text{d}V.
\end{aligned}
\label{eq:energy_balance}
\end{equation}
The terms on the right-hand side represent heat flux, mass advection, compressional work, electron-ion equilibration, and radiative emission and absorption, respectively.

Because the shocked plasma volume evolves in time, with the Mach-number-truncated region defining time-dependent boundary, the Reynolds transport theorem~\cite{reynolds_1903} must be applied:
\begin{equation}
\frac{\text{d}}{\text{d}t} \int p_{\text{th},e}~\text{d}V = \int \frac{\partial p_{\text{th},e}}{\partial t} ~\text{d}V + \int p_{\text{th},e}\boldsymbol{v} \cdot \hat{\boldsymbol{n}}~\text{d}A,
\label{eq:reynolds_transport_theorem}
\end{equation}
where $\boldsymbol{v}$ is the velocity of the evolving truncated boundary.

\begin{figure*}
    \centering
    \includegraphics[width=0.75\linewidth]{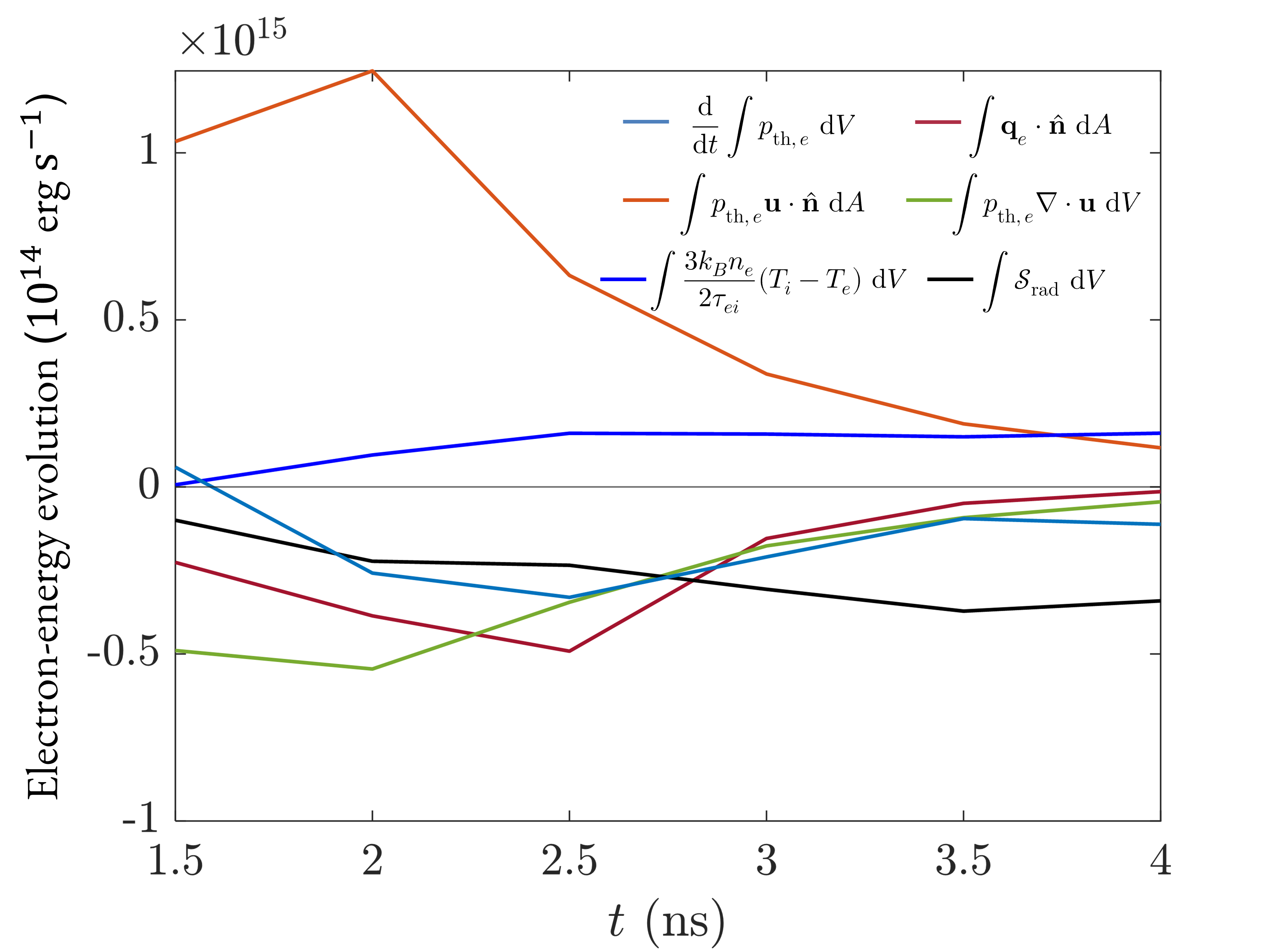}
    \caption{\textbf{Electron-energy balance.} Evolution of internal electron energy within shocked plasma. The pale blue line shows the total energy change, the red shows the heat-flux losses, the orange line shows the rate of total mass advection, green is the expansion losses, dark blue is the ion-electron equilibration, black shows the radiative emission losses, and the grey indicates the zero-line.}
    \label{fig:energy_balance}
\end{figure*}

Extended Data Fig.~\ref{fig:energy_balance}a shows the evolution of the internal electron energy within the shocked plasma, together with each term from Eq.~(\ref{eq:energy_balance}). When evaluating the heat-flux integral, heat flux in the $z$ direction was neglected. Owing to the large Hall parameter, cross-field transport is expected to be strongly suppressed, as discussed in the main text. The heat flux was also simulated using a Larsen flux limiter, capped at $6\%$ of the free-streaming limit; further details of the heat-flux model are given in Ref.~\cite{tzeferacos_2015}.

For the radiative source and sink term, radiative energy gains were assumed to be negligible. This assumption is justified because the shocked plasma is optically thin to its own spectral-line emission, while inverse bremsstrahlung absorption is comparatively small, $\mathcal{O}(10^{12}~\text{erg s}^{-1})$, relative to the other terms. To calculate the radiative losses, the FLASH output was processed using SPECT3D to generate a spatially resolved self-emission map of the plasma, following the approach of Ref.~\cite{vincent_2026}. The shocked-plasma region was then spatially averaged to estimate the emissivity as a function of time.

The remaining terms were calculated directly from the FLASH domain without additional assumptions. 

\section*{Acknowledgements}
The authors would like to thank Henry Edwards at the Central Laser Facility for their contribution to target fabrication.

\section*{Author contributions}
The experiment was carried out by T.A.V., A.F.A.B, T.H., S.P., P.Th., and L.C., with support from G.G. Target fabrication was carried out by P.A. and C.S. The data analysis was carried out by T.A.V. The paper was written by T.A.V. and A.F.A.B., with further contributions from all coauthors. FLASH simulations were carried out by P.Tz., and analysed by T.A.V. Further experimental, simulation, and theoretical support was provided by C.D., R.D., D.L., J.M., C.A.J.P., H.P., E.R.T., L.W., and W.G. This project was conceived by A.F.A.B. and P.Tz.

\section*{Funding}
The research leading to these results has received funding from AWE plc, UKRI (grant numbers MR/W006723/1, EP/Y035038/1), STFC (grant number ST/W000903/1), U.S. DOE NNSA (awards DE-NA0002724, DE-NA0003605, DE-NA0003934, DE-NA0004144, DE-NA0004147; subcontracts 630138 and C4574 with LANL), U.S. DOE Office of Science (award DE-SC0021990), the NSF (awards PHY-2033925 and PHY-2308844), and the EPSRC and First Light Fusion under the AMPLIFI Prosperity Partnership (EP/X025373/1). The software used in this work was developed in part by the U.S. DOE NNSA and U.S. DOE Office of Science-supported Flash Center for Computational Science. UK Ministry of Defence © Crown owned copyright 2026/AWE.

\section*{Competing interests}
The authors declare no competing interests.

\section*{}

\begin{table*}[h]
\begin{center}
  \renewcommand{\arraystretch}{1.5}
\def~{\hphantom{0}}
  \begin{tabular}{p{0.35\textwidth}cc}

         Quantity & Formula & Value\\
         \hline
         Electron temperature ($T_{e}$) & - & $1.1~\text{keV}$ \\
         Electron number density ($n_{e}$) & - & $1\times10^{20}~\text{cm}^{-3}$ \\
         RMS peak magnetic field strength ($B$) & - & $210~\text{kG}$ \\
         
         Ablation plume flow velocity ($u_{\text{plume}}$) & - & $9\times 10^7~\text{cm}~\text{s}^{-1}$\\
        
         
        Coulomb logarithm ($\log\Lambda$) & $\begin{aligned} 23.5 &- \log n_{e}^{1/2}T_{e}^{-5/4}\\ 
        & - \sqrt{1\times10^{-5} + (\log(T_{e}) - 2)^{2}/16} \end{aligned}$ & 8 \\
         
        Electron mean-free path ($\lambda_{e}$) & $2.1\times10^{13}T_{e}^{2}/Z_{\text{eff}}n_{e}\log\Lambda$ & $40~\mu\text{m}$ \\
         
        Electron gyroradius ($\rho_{e}$) & $2.4T_{e}^{1/2}|\boldsymbol{{B}}|^{-1}$ & $4~\mu\text{m}$\\
        
        Temperature gradient length scale ($L_{T}$) & $\nabla_{||}\log T_{e}$ & $600~\mu\text{m}$\\
        
        Electron plasma beta ($\beta_{e}$) & $4\times10^{-11}n_{e}T_{e}|\boldsymbol{{B}}|^{-2}$ & $100$\\

        Magnetic resistivity ($\eta$) & $3.1\times10^{5}Z_{\text{eff}}\log\Lambda/T_{e}^{3/2}$ & $300~\text{cm}^{2}~\text{s}^{-1}$\\

        Magnetic Reynolds number ($\text{Rm}$) & $u_{\text{plume}}L_{s}/\eta$ & $2\times10^{4}$\\
        
        Hall parameter ($\text{Ha}_{e}$) & $\lambda_{e}/\rho_{e}$ & $11$\\

        Spitzer-H\"arm thermal conductivity ($\kappa_{\mathrm{\text{S}}}$) & $5.2\times10^{21} T_{e}^{5/2}/Z_{\rm eff} \log\Lambda$  & $3 \times 10^{27}~\text{cm}^{-1}~\text{s}^{-1}$ \\
        
        Yerger effective parallel thermal conductivity ($\kappa_{\mathrm{\text{Y}}}/\kappa_{\mathrm{\text{S}}}$) & $0.8[\beta_{e}\lambda_{e}/L_{T}]^{-1}\kappa_{\text{S}}$ & $0.05$ \\
         
        Ryutov effective parallel thermal conductivity ($\kappa_{\mathrm{R}}/\kappa_{\mathrm{\text{S}}}$) & $ (\rho_{e}/\lambda_{e})\kappa_{\text{S}}$ & $0.09$\\
         
         
         
         \hline
         \\
  \end{tabular}
  \caption{\textbf{Summary of key plasma parameters used throughout analysis.} Numerical coefficients used above are the following, for a parylene-C plasma: effective ion charge $Z_{\text{eff}} = 7.7$, $Z_{\text{q,eff}} = 4.44$ using quasineutrality, average atomic weight $m_{\text{eff}} = 7.80$, and adiabatic index $\gamma = 5/3$. The parameter expressions can be found in ~\cite{bott_pnas_2021, vincent_2026}. The stated values are calculated using the $2.2~\text{ns}$ data quoted in the main text. Formulae in CGS units with temperature in eV}
  \label{table:equations_for_calcs}

\end{center}
\end{table*}

\end{document}